\documentclass[aps,nofootinbib,amsfonts,superscriptaddress]{revtex4}
\usepackage{amsfonts,amssymb,amscd,amsmath}
\usepackage{graphicx}
\usepackage{subfigure}
\usepackage{bm}
\graphicspath{{figures/}}

\newcommand{\dd}{\, \mathrm{d}}

\begin{document}

\begin{flushright}
LU TP 15-37\\
October 2015
\vskip1cm
\end{flushright}

\title{Drell-Yan phenomenology in the color dipole picture revisited}

\author{Eduardo Basso}
\email{eduardo.basso@thep.lu.se}
\affiliation{Instituto de F\'{\i}sica, Universidade Federal do Rio de Janeiro, Caixa Postal 68528, Rio de Janeiro, RJ 21941-972, Brazil}
\affiliation{Department of Astronomy and Theoretical Physics, Lund University, SE-223 62 Lund, Sweden} 

\author{Victor P.  Goncalves}
\email{victorpbg@thep.lu.se}
\affiliation{Department of Astronomy and Theoretical Physics, Lund University, SE-223 62 Lund, Sweden} 
\affiliation{High and Medium Energy Group, Instituto de F\'{\i}sica e Matem\'atica, 
Universidade Federal de Pelotas, Pelotas, RS, 96010-900, Brazil} 

\author{Jan Nemchik}
\email{nemcik@saske.sk}
\affiliation{Czech Technical University in Prague, FNSPE, B\v rehov\'a 7, 11519 Prague, Czech Republic}
\affiliation{Institute of Experimental Physics SAS, Watsonova 47, 04001 Ko\v sice, Slovakia}

\author{Roman Pasechnik}
\email{roman.pasechnik@thep.lu.se}
\affiliation{Department of Astronomy and Theoretical Physics, Lund University, SE-223 62 Lund, Sweden}

\author{Michal \v{S}umbera}
\email{sumbera@ujf.cas.cz}
\affiliation{Nuclear Physics Institute ASCR, 25068 \v{R}e\v{z}, Czech Republic}

\begin{abstract}
An extensive phenomenological study of the Drell-Yan (DY) process in $pp$ collisions at various energies is performed in the color dipole framework. Besides previously studied 
$\gamma^*$ production we have also included the $Z^0$ contribution relevant at large dilepton invariant masses. We investigate the DY cross section differential in invariant mass, rapidity and transverse momentum of the dilepton pair in $pp$ collisions at RHIC and LHC. We consider three different phenomenological models for the dipole cross section and found a reasonable agreement with the available data. As a further test of the color dipole formalism, we also study the correlation function in azimuthal angle between the dilepton pair and a forward pion $\Delta\phi$ for different energies, dilepton rapidites and invariant masses. The characteristic double-peak structure of the correlation function around $\Delta \phi\simeq \pi$ found for very forward pions and low-mass dilepton pairs is sensitive to the saturation effects and can be tested by future DY measurements in $pp$ collisions.
\end{abstract}
\maketitle

\section{Introduction}

The study of Drell-Yan (DY) processes at LHC energies provide an important test 
of the Standard Model (SM) as well as can supply with an additional information about 
New Physics beyond the SM. In particular, the DY process in $pp/pA/AA$ collisions at the LHC 
is an excellent tool for the investigations of strong interaction dynamics in an extended 
kinematical range of energies and rapidities (for a recent review see, e.g. Ref.~\cite{Peng}). 
For example, recent measurements of the gauge boson production cross section by the LHCb 
experiment \cite{lhcb_data} at forward rapidities have a sensitivity to $x$ values down to 
$1.7 \times 10^{-4}$ at the scale $Q^2 \sim M^2$ ($M$ is the invariant 
mass of the dilepton) probing the parton distribution functions (PDFs) as well as 
soft QCD dynamics and non-linear effects in a kinematical range different from that studied by HERA.

During the last two decades several approaches have been proposed to improve the fixed-order 
QCD perturbation theory description of the DY process which is not reliable when two or more different 
hard scales are present (see e.g. Refs.~\cite{Collins,nnlo,Qiu,Qiu2,wmr,baranov,deak,hautmann,ball,dynnlo,Bonvini}). 
A well-known example is the description of the transverse momentum $p_T$ distribution of the dilepton. 
In the low-$p_T$ region, $p_T\ll M$, there are two powers of $\ln(M^2/p_T^2)\gg 1$ 
for each additional power of the strong coupling $\alpha_s$, and the DY $p_T$ distribution calculated in 
fixed-order QCD perturbation theory is not reliable. Only after resummation of the large terms
$\propto \alpha_s^n\ln^{2n+1}(M^2/p_T^2)$ the predictions become consistent 
with the data. Another example is in the case of the high energies $s\gg M^2$ when 
potentially large terms $\propto\alpha_s^n\ln^n(s/M^2)$ should also be resummed. In this case, 
the standard collinear factorisation approach should be generalized by taking into account the transverse 
momentum evolution of the incoming partons and QCD nonlinear effects.

\begin{figure}[t]
\centering
\subfigure[]{
\scalebox{0.27}{\includegraphics{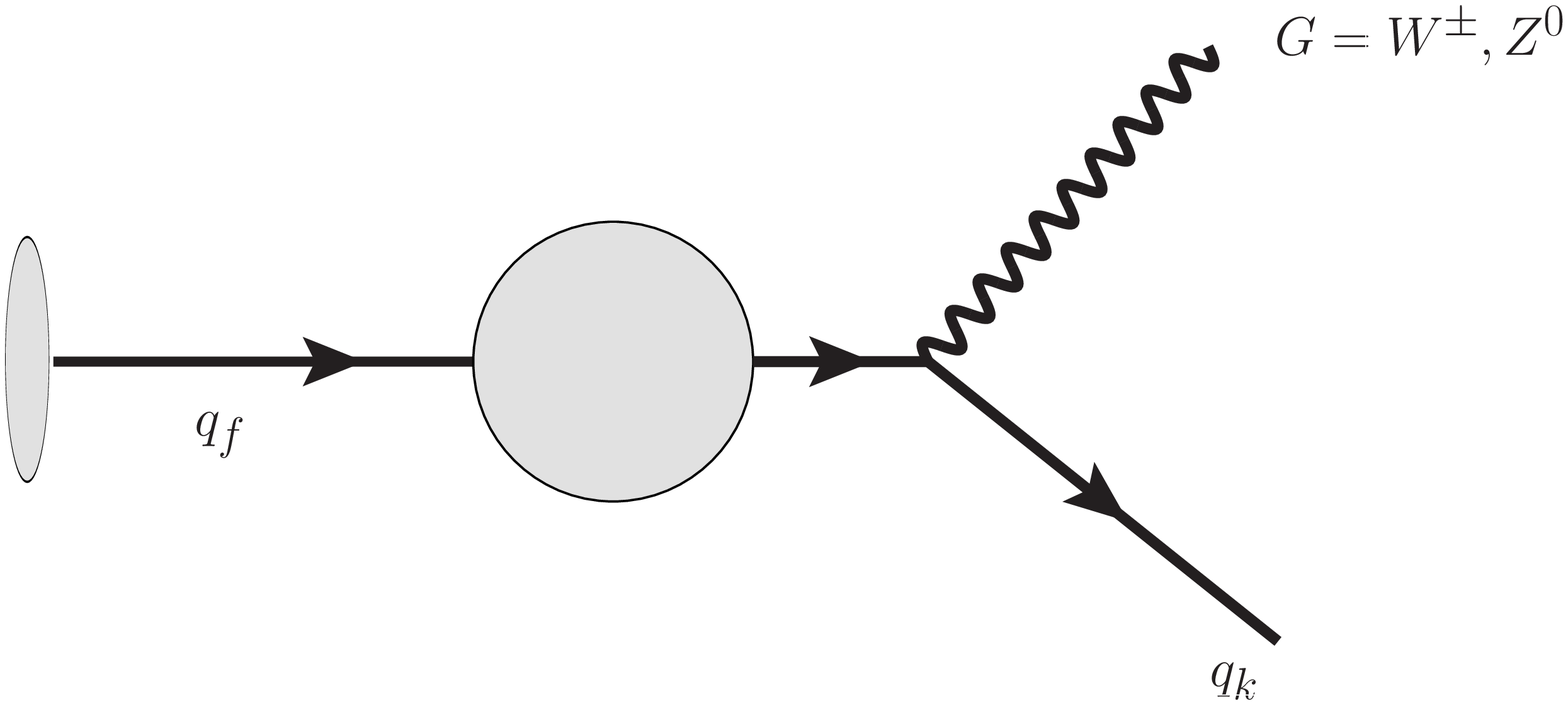}}
\label{fig:gb_dir}
}
\centering
\subfigure[]{
\scalebox{0.27}{\includegraphics{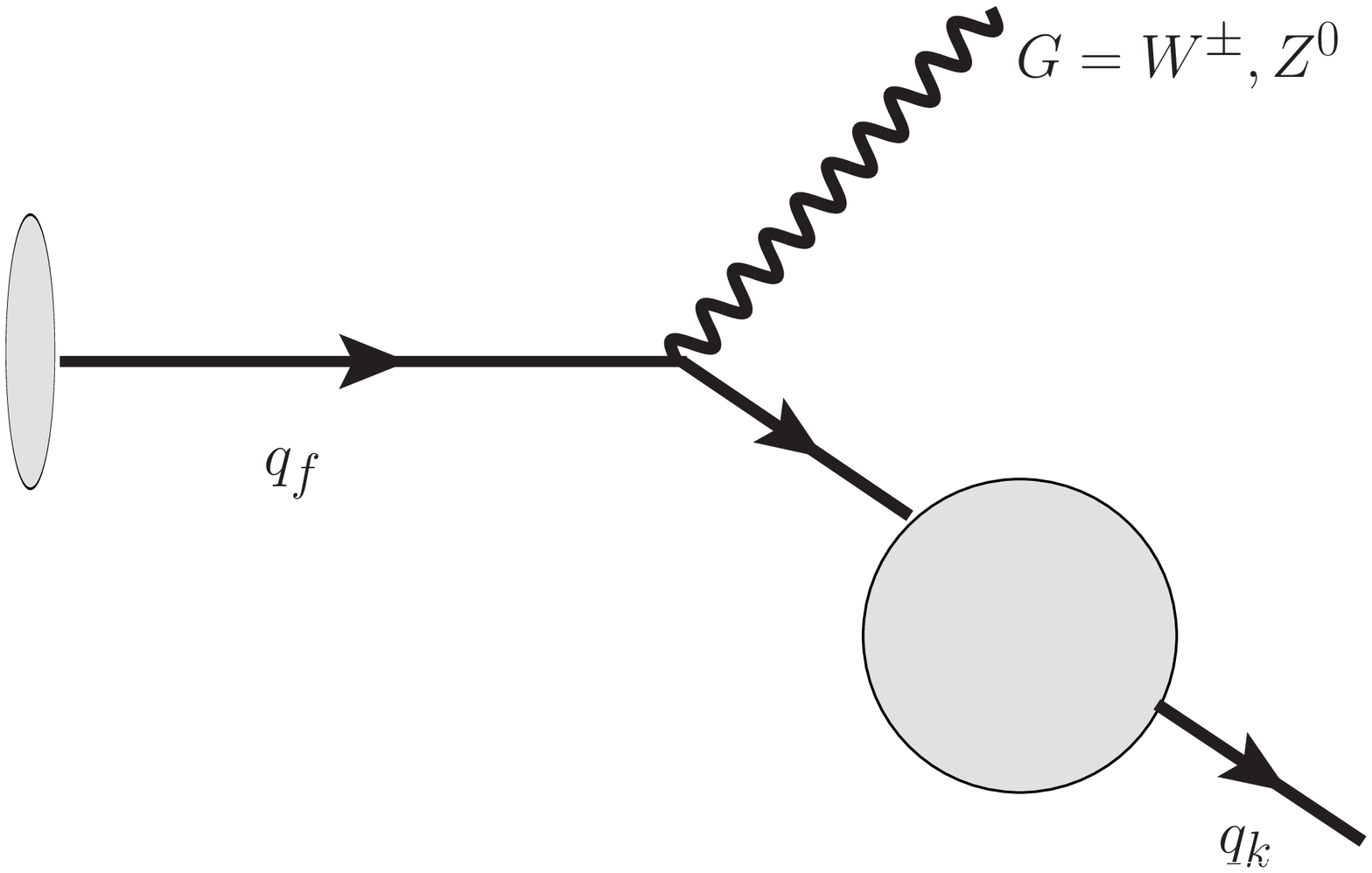}}
\label{fig:gb_frag}
}
\centering
\subfigure[]{
\scalebox{0.27}{\includegraphics{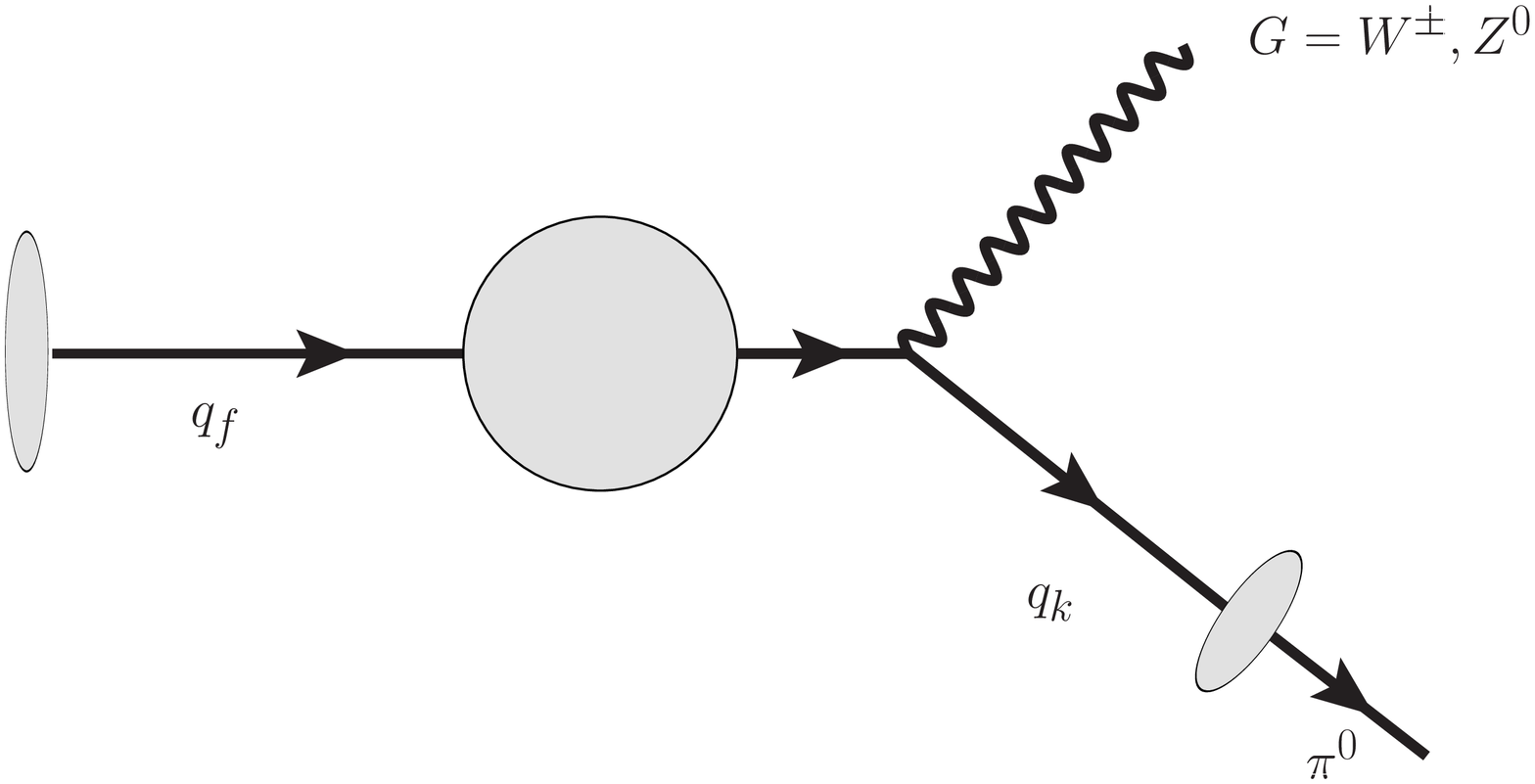}}
\label{fig:gb_frag1}
}
\caption{Diagrams (a) and (b) represent the process of a gauge boson radiation by a quark (antiquark) 
of flavour $f$ either after or before the interaction with the target color field (denoted by a shaded circle), 
respectively. For the considered $\gamma,\,Z^0$ radiation $q_k = q_f$. Diagram (c) represents the gauge 
boson-pion production in the color dipole picture.}
\label{fig:gb_dip}
\end{figure}

One of the phenomenological approaches which effectively takes into account the higher-order QCD corrections 
is the color dipole formalism \cite{nik}. At high energies, color dipoles with a definite transverse separation are 
eigenstates of interaction. The main ingredient of this formalism is the dipole-target scattering cross section 
which is universal and process-independent and thus can be determined phenomenologically, for example, from 
the Deep Inelastic Scattering (DIS) data \cite{GBW}. In particular, it provides a unified description of inclusive and diffractive 
observables in $ep$ scattering processes as well as other processes in hadron-hadron collisions such 
as DY, prompt photon, heavy quark production etc \cite{nik,nik_dif,k95,bhq97,kst99,krt01,npz}. 
Although cross sections are Lorentz invariant, the partonic interpretation of the corresponding 
processes depends on the reference frame \cite{k95}. In particular, in the framework of 
conventional parton model the DY process is typically considered as due to parton annihilation 
in the center-of-mass frame description. In the target rest frame the same process can be viewed 
as a bremsstrahlung of $\gamma/Z^0$ in the dipole picture as is illustrated in Figs.~\ref{fig:gb_dip} (a) and (b). 
In the latter case, the radiation occurs both after and before the quark scatters off the target and the corresponding
amplitudes interfere. In the high-energy limit, the projectile quark probes dense gluonic field in the target such
that nonlinear effects due to multiple scatterings become important and should be taken into account.

The DY process mediated by virtual photon has been studied within the dipole framework in the literature by several authors 
(see e.g. Refs.~\cite{dynuc,rauf,gay}). In particular, in Ref.~\cite{rauf} it has been demonstrated that the dipole model
provides as precise prediction for the DY cross section as the NLO collinear factorisation framework giving a solid foundation
for the current more extensive study. The inclusive gauge bosons production has been previously analysed by some of 
the authors in Ref.~\cite{Basso} where predictions for the total cross sections and rapidity distributions were found 
to be in a good agreement with the recent LHC data. In the diffractive channel, the DY and electroweak gauge boson
production processes have been studied in the dipole framework in Ref.~\cite{pkp}.

The goal of the current work is the following. First, we update and improve previous studies. We present predictions for 
the transverse momentum, invariant mass and rapidity distributions for the DY pair production at RHIC and LHC energies 
and compare them with available data taking into account $Z^0$ boson contribution in addition to the virtual photon. 
Second, we present a detailed analysis of the azimuthal correlation between the DY pair and a forward pion (see Fig.~\ref{fig:gb_dip}(c)). 
Similar correlations in dihadron, real photon-hadron and dilepton-hadron channels have been previously investigated in 
Refs.~\cite{Marquet,stasto,stastody,amir}. In variance from the dihadron channel, the dilepton-hadron correlations can serve
as an efficient probe of the initial state effects since the intermediate virtual boson $(\gamma/Z^0)$ does not interact 
with partons inside the target hadron and therefore the final state interaction effects do not exist. In this paper, for 
the first time we present results for such an observable in $pp$ collisions at RHIC ($\sqrt{s} = 200$ and 500 GeV) and 
LHC  ($\sqrt{s} = 7$ and 14 TeV) at different $M$. We test three different models for the dipole cross section accounting 
for saturation effects \cite{hdqcd} in order to estimate the underlined theoretical uncertainties.

This paper is organized as follows. In the next Section, we present a brief overview of gauge boson production in the color dipole framework. 
Moreover, we derive the differential cross section for the dilepton-hadron production in the momentum representation taking into account 
both virtual photon and $Z^0$ boson contributions. In Section \ref{res}, we present our results for the total cross sections, invariant mass, 
rapidity and transverse momentum distributions and compare our predictions with the available data at different energies. Predictions for 
future RHIC and LHC runs are also given. Furthermore, the azimuthal correlation function is evaluated for the DY-pion production in $pp$ 
collisions at RHIC and LHC for different dilepton invariant masses and rapidities. For the first time, we have found a double-peak structure 
in the pion-dilepton correlation function around $\Delta \phi = \pi$ at forward pion rapidities. Finally, in Section \ref{conc}, our main conclusions 
are summarized.

\section{Inclusive gauge boson production in the dipole picture}
\label{formalism}

As was mentioned above, in the color dipole picture the DY process is considered as a bremsstrahlung of a virtual gauge boson $G^*$ 
by a projectile quark, where $G=\gamma,Z^0,W^\pm$ \cite{k95,bhq97,kst99} as is illustrated in Fig.~\ref{fig:gb_dip}. In the high 
energy limit, each of the two graphs factorizes into the production vertex for a given gauge boson times the scattering amplitude of a quark off 
the target. The quark scatters at different impact parameters depending on whether the gauge boson is radiated after 
or before the scattering. The interference between these scattering amplitudes implies that the squared matrix element 
for the gauge boson production is expressed in terms of the universal dipole-target cross section $\sigma_{q\bar{q}}(\rho,x)$ 
with transverse separation between $\rho$ initial $(q_f)$ and final $(q_k)$ quark as is shown in Fig.~\ref{fig:gb_dip}.

\subsection{Inclusive DY cross section}

In order to estimate the hadronic cross section for the inclusive DY process $pp\rightarrow G^* X$ one has to note 
that the gauge boson carries away the light-cone momentum fraction $x_1$ and $\alpha$ from the projectile proton 
and quark emitting the gauge boson, respectively. Consequently, the momentum fraction of the quark is given 
by $x_q = x_1/\alpha$. Then the cross section for the inclusive gauge boson production with invariant mass $M$
and transverse momentum $p_T$ is expressed in terms of the quark (antiquark) densities $q_f$ ($\bar{q}_f$)
at momentum fraction $x_q$ as follows
\begin{eqnarray} \label{eq:gb_cs}
\frac{d \sigma (pp\rightarrow G^* X)}{\dd^2 p_T d\eta} = J(\eta, p_T)\,\frac{x_1}{x_1 + x_2}\,
\sum_f\sum_{\lambda_G=L,T}  \int_{x_1}^1 \frac{d \alpha}{\alpha^2} \left[ q_f(x_1/\alpha,\mu_F^2) + \bar{q}_{{f}}(x_1/\alpha,\mu_F^2)  \right] 
\frac{d\sigma^f_{\lambda_G} (qN \rightarrow qG^*X)}{d\ln \alpha d^2p_T}
\end{eqnarray}
where 
\begin{equation}
J(\eta, p_T)\equiv \frac{dx_F}{d\eta} = \frac{2}{\sqrt{s}} \sqrt{M^2 + p_T^2}\, \cosh(\eta)
\end{equation}
is the Jacobian of transformation between Feynman variable $x_F = x_1 - x_2$ and pseudorapidity $\eta$ of the virtual gauge boson 
$G^*$ and $\mu_F^2=p_T^2+(1-x_1)M^2$ is the factorization scale in quark PDFs. In practical calculations below we take $\mu_F\simeq M$, 
for simplicity. We have checked numerically that such a choice of the factorisation scale is a good approximation in the whole kinematical range we are
concerned about in this work. The dilepton cross section analysed below is related to the inclusive $G=\gamma,\,Z^0$ production cross section (\ref{eq:gb_cs}) 
as follows
\begin{eqnarray} \label{dilep}
\frac{d \sigma (pp\rightarrow [G^*\to l\bar l] X)}{d^2 p_T dM^2 d\eta}=
{\cal F}_G(M)\,\frac{d \sigma (pp\rightarrow G^* X)}{d^2 p_T d\eta} \,,
\end{eqnarray}
where
\begin{eqnarray}
{\cal F}_\gamma(M)=\frac{\alpha_{em}}{3\pi M^2} \,, \qquad  {\cal F}_Z(M)=\mathrm{Br}(Z^0\to l\bar l)\rho_Z(M) \,.
\end{eqnarray}
Here, the branching ratio $\mathrm{Br}(Z^0\to l\bar l)\simeq 0.101$, and $\rho_Z(M)$ is the invariant mass distribution of the $Z^0$ boson
in the narrow width approximation
\begin{eqnarray}
\rho_Z(M)=\frac{1}{\pi}\, \frac{M\Gamma_Z(M)}{(M^2-m_Z^2)^2+[M\Gamma_Z(M)]^2}\,, \qquad \Gamma_Z(M)/M\ll 1 \,,
\end{eqnarray}
in terms of the on-shell $Z^0$ boson mass, $m_Z\simeq 91.2$ GeV, and the generalized total $Z^0$ decay width
\begin{eqnarray}
\Gamma_Z(M)=\frac{\alpha_{em}M}{6\sin^22\theta_W}\Big(\frac{160}{3}\sin^4\theta_W - 40 \sin^2\theta_W + 21\Big) \,,
\end{eqnarray}
where $\theta_W$ is the Weinberg gauge boson mixing angle in the SM, $\sin^2\theta_W\simeq 0.23$, and 
$\alpha_{em}=e^2/(4\pi)= 1/137$ is the fine structure constant.

The transverse momentum distribution of the gauge boson $G^*$ can be obtained by a generalization of the well-known formulas for the photon 
bremsstrahlung \cite{dynuc,rauf}. The corresponding differential cross section for a given incoming quark flavour $f$ reads
\begin{eqnarray}
\label{ptdistcc}
\frac{d\sigma^f_{T,L} (qN \rightarrow qG^*X)}{d \ln \alpha d^2p_T} & = & \frac{1}{(2\pi)^2}\, \sum_\text{quark pol.}
\int d^2\rho_1 d^2\rho_2 \exp[i{\bf p}_T \cdot ({\bm\rho}_1 - {\bm\rho}_2)]\, \Psi^{V - A}_{T,L}(\alpha,{\bm\rho}_1,m_f) 
\Psi^{V - A, *}_{T,L}(\alpha,{\bm\rho}_2,m_f) \nonumber \\
 & \times & \frac{1}{2}\left[ \sigma_{q\bar{q}}(\alpha {\bm\rho}_1,x_2) + \sigma_{q\bar{q}}(\alpha {\bm\rho}_2,x_2) - 
 \sigma_{q\bar{q}}(\alpha|{\bm\rho}_1- {\bm\rho}_2|,x_2)\right]\,, 
\end{eqnarray}
where $x_2 = x_1 - x_F$, ${\bm\rho}_1$ and ${\bm\rho}_2$ are the quark-$G$ transverse separations in the total radiation 
amplitude and its conjugated counterpart, respectively. Assuming that the projectile quark is unpolarized, the products of the 
vector and axial-vector wave functions in Eq.~(\ref{ptdistcc}) can be written as follows
\begin{eqnarray} \label{Psi2}
&  & \sum_\text{quark pol.} \Psi^{V-A}_{T,L}(\alpha,{\bm\rho}_1,m_f) \Psi^{V-A,*}_{T,L}(\alpha,{\bm\rho}_2,m_f)  = \nonumber \\ 
& = & \Psi^V_{T,L}(\alpha,{\bm\rho}_1,m_f)\Psi^{V,*}_{T,L}(\alpha,{\bm\rho}_2,m_f) + \Psi^A_{T,L}(\alpha,{\bm\rho}_1,m_f) 
\Psi^{A,*}_{T,L}(\alpha,{\bm\rho}_2,m_f)\,,
\end{eqnarray}
where the averaging over the initial and summation over final quark helicities is performed and the quark flavour dependence comes 
only via projectile quark mass $m_f$. Different components in Eq.~(\ref{Psi2}) read \cite{pkp}
\begin{eqnarray}\label{VV}
&&\Psi^{T}_{V}\Psi^{T*}_{V}= \frac{({\cal C}^G_f)^2(g^{G}_{v,f})^2}{2\pi^2}\Bigg\{
     m_f^2 \alpha^4 {\rm K}_0\left(\tau \rho_1\right)
     {\rm K}_0\left(\tau \rho_2\right)+ \left[1+\left(1-\alpha\right)^2\right]\tau^2
   \frac{{\bm\rho}_1\cdot{\bm\rho}_2}{\rho_1 \rho_2}
     {\rm K}_1\left(\tau \rho_1\right)
     {\rm K}_1\left(\tau \rho_2\right)\Bigg\}\,,\nonumber \\ 
&&\Psi^{L}_{V}\Psi^{L*}_{V}=
\frac{({\cal C}^G_f)^2(g_{v,f}^G)^2}{\pi^2}M^2
\left(1-\alpha\right)^2
  {\rm K}_0\left(\tau \rho_1\right)
     {\rm K}_0\left(\tau \rho_2\right)\,, \nonumber \\
&&\Psi^{T}_{A}\Psi^{T*}_{A}=
   \frac{({\cal C}^G_f)^2(g_{a,f}^G)^2}{2\pi^2}\Bigg\{
     m_f^2 \alpha^2(2-\alpha)^2 {\rm K}_0\left(\tau \rho_1\right)
     {\rm K}_0\left(\tau \rho_2\right)+ \left[1+\left(1-\alpha\right)^2\right]\tau^2
   \frac{{\bm\rho}_1\cdot{\bm\rho}_2}{\rho_1 \rho_2}
     {\rm K}_1\left(\tau \rho_1\right)
     {\rm K}_1\left(\tau \rho_2\right)\Bigg\}\,, \nonumber \\
&&\Psi^{L}_{A}
\Psi^{L*}_{A} = \frac{({\cal C}^G_f)^2(g_{a,f}^G)^2}{\pi^2}\frac{\tau^2}{M^2}\Bigg\{\tau^2
  {\rm K}_0\left(\tau \rho_1\right)
     {\rm K}_0\left(\tau \rho_2\right)+\alpha^2m_f^2\frac{{\bm\rho}_1\cdot{\bm\rho}_2}{\rho_1 \rho_2}
     {\rm K}_1\left(\tau \rho_1\right)
     {\rm K}_1\left(\tau \rho_2\right)\Bigg\}\,, 
\end{eqnarray}
where $\tau^2 = (1-\alpha)M^2 + \alpha^2 m_f^2$, ${\rm K}_{0,1}$ denote the modified Bessel functions 
of the second kind, and the coupling factors ${\cal C}^G_f$ are defined as
\begin{eqnarray}
 {\cal C}^{\gamma}_f=\sqrt{\alpha_{em}} e_f\,,\qquad {\cal C}^Z_f=\frac{\sqrt{\alpha_{em}}}{\sin 2\theta_W}\,,\qquad 
 {\cal C}^{W^+}_f=\frac{\sqrt{\alpha_{em}}}{2\sqrt{2}\sin\theta_W}V_{f_uf_d}\,,\qquad
{\cal C}^{W^-}_f=\frac{\sqrt{\alpha_{em}}}{2\sqrt{2}\sin\theta_W}V_{f_df_u}\,, \label{Cfacs}
\end{eqnarray}
with the vectorial coupling at the leading order given by
\begin{eqnarray}
g_{v,f_u}^Z=\frac12-\frac43\sin^2\theta_W\,,\qquad
g_{v,f_d}^Z=-\frac12+\frac23\sin^2\theta_W\,,\qquad g_{v,f}^W=1\,, \label{vec}
\end{eqnarray}
and 
\begin{eqnarray}
g_{a,f_u}^Z=\frac12\,,\qquad g_{a,f_d}^Z=-\frac12\,,\qquad g_{a,f}^W=1 \label{axial}
\end{eqnarray}
in the axial-vector case. In the above formulas, $f_u=u,c,t$ and $f_d=d,s,b$ are the flavours of up- and down-type quarks, respectively, 
$V_{f_uf_d}$ is the CKM matrix element corresponding to $f_u\to f_d$ transition, and $e_f$ is the charge of the projectile quark\footnote{Also, 
we are focused only on light quark flavours $f=u,d,s$}. In the case of projectile photon we have $g_{v,q} = 1$ and $g_{a,q} = 0$. 
In the present analysis, we restrict ourselves to study of dilepton $l\bar l$ productions channels in $pp$ collisions and therefore 
we consider production of virtual $\gamma$ and $Z^0$ bosons only. For this reason, we leave $W^\pm$  production in 
the $l\bar\nu_l$ decay channel for future studies. Integrating over the phase space of the final quark, Eqs.~(\ref{eq:gb_cs}) and (\ref{dilep}) 
enable us to study the (pseudo)rapidity, transverse momentum and invariant mass distributions for the DY process. 

\subsection{Dilepton-hadron azimuthal correlations}

In order to study the azimuthal angle correlation between the DY pair and a hadron in the final state, we should keep 
an information about the quark which radiates the virtual gauge boson $G^*$. This analysis can, in principle, be carried out 
in the impact parameter representation (for more details, see Appendix B in Ref.~\cite{dynuc} for the $\gamma^*$ case), but numerically 
it is rather cumbersome due to a large number of oscillating Fourier integrals. To avoid this complication we switch to the derivation
of the corresponding differential cross section in momentum representation as was performed in the $\gamma^*$ case in 
Refs.~\cite{jamal,amir,Dominguezetal12,tese}. Additionally, we extend it by incorporating $Z^0$ boson contribution relevant 
at large dilepton invariant masses. Our basic goal here is to investigate the dilepton-pion correlations accounting for both 
virtual $\gamma$ and $Z^0$ contributions in $pp$ collisions at high energies and their interference in various kinematical 
domains in rapidity and dilepton invariant mass. The latter can be straightforwardly generalized to the proton-nucleus case.

A generalisation of the results in Refs.~\cite{Dominguezetal12,tese} is achieved by accounting for both vector and axial contributions in
the gauge boson distribution amplitude $q\to q+G^*$ with unpolarised $q$ and $G^*$. This leads to the differential cross section 
for the production of a virtual gauge boson $G^*$ and a hadron $h$ (for simplicity, we take $m_f = 0$ for $f=u,d,s$ in what follows)
\begin{eqnarray}
\frac{d \sigma(pp \to h G^* X)}{d Y d y_h d^2p_T d^2p_T^h } & = & 
\int_{\frac{x_h}{1-x_1}}^1 \frac{\dd z_h}{z_h^2} \sum_f D_{h/f}(z_h,\mu_F^2)\, 
x_p q_f(x_p,\mu_F)\, (1-z) S_{\perp}\, F(x_g, k^g_T) \nonumber\\
& \times & \left\{  \frac{({\cal C}^G_f)^2 g_{v,f}^2}{2\pi} \left[  \left( 1 + (1-z)^2 \right) \frac{z^2 {k^g_T}^2}
{\left[ P_T^2 + \epsilon_M^2  \right] \left[ ({\bf P}_T + z{\bf k}^g_T )^2 + \epsilon_M^2 \right]} \right.\right.\nonumber\\
& \quad & \qquad \qquad \left.\left. -\; z^2\epsilon_M^2 \left( \frac{1}{P_T^2 + \epsilon_M^2} - \frac{1}
{({\bf P}_T + z{\bf k}^g_T)^2 + \epsilon_M^2} \right)^2 \right] \right.\nonumber \\
& \quad & \left. +\;  \frac{({\cal C}^G_f)^2 g_{a,f}^2}{2\pi} \left[  \left( 1 + (1-z)^2 \right) \frac{z^2 {k^g_T}^2}
{\left[ P_T^2 + \epsilon_M^2  \right] \left[ ({\bf P}_T + z{\bf k}^g_T)^2 + \epsilon_M^2 \right]} \right.\right.\nonumber\\
& \quad & \qquad \qquad \left.\left.  - \frac{z^2 \epsilon_M^4}{M^2} \left(\frac{1}{P_T^2 + \epsilon_M^2} - 
\frac{1}{({\bf P}_T + z{\bf k}^g_T)^2 + \epsilon_M^2} \right)^2 \right] \right\}\,,\label{eq:dy-dijet}
\end{eqnarray}
where the couplings ${\cal C}^G_f$ and $g_{v/a,f}$ are given in Eqs.~(\ref{Cfacs}) -- (\ref{axial}), $D_{h/f}$ is the 
fragmentation function of the projectile quark $q$, which has emitted the gauge boson $G^*$, into the produced 
hadron $h$. In addition, in Eq.~(\ref{eq:dy-dijet}) variables $Y$ (${\bf p}_T$) and $y_h$ (${\bf p}_T^h$) are the rapidities (transverse momenta) 
of the gauge boson $G^*$ and the hadron $h$ in the final state, respectively, $z_h$ is the momentum fraction of the hadron $h$ relative 
to the quark $q$ it fragments from, and $S_{\perp}$ is the transverse area of the target whose explicit form is not needed for our purposes here.
Other kinematics variables are defined as follows
\begin{eqnarray}
&& x_1 = \sqrt{\frac{p_T^2 + M^2}{s}}\,e^{Y} \,, \qquad x_h \simeq \frac{p_T^h}{\sqrt{s}}\,e^{y_h} \,, \qquad 
x_p = x_1 + \frac{x_h}{z_h} \,, \qquad z = \frac{x_1}{x_p} \,, \qquad \epsilon_M^2= (1-z) M^2 \,, \\ 
&& x_g = x_1\,e^{ - 2 Y} +  \frac{x_h}{z_h}\,e^{ - 2 y_h} \,, \qquad 
{\bf k}^q_T = \frac{{\bf p}_T^h}{z_h} \,, \qquad {\bf k}^g_T = {\bf p}_T + {\bf k}^q_T \,, \qquad
{\bf P}_T = (1-z){\bf p}_T - z{\bf k}^q_T \,,
\end{eqnarray}
where $x_1$ and $x_h$ are the gauge boson $G$ and the hadron $h$ momentum fractions taken from the incoming proton,
${\bf P}_T$ is the relative transverse momentum between the gauge boson $G^*$ and the quark $q$, ${\bf k}^q_T$ is the transverse
momentum of the quark $q$ in the final state, ${\bf k}^g_T$ is the transverse momentum of the exchanged gluon\footnote{Variables
$z$ and $x_p$ have the same physical meaning as $\alpha$ and $x_q\equiv x_1/\alpha$ in Eq.~(\ref{eq:gb_cs}), respectively, but
now they are related to the kinematic variables corresponding the final hadron $z_h$, $y_h$ and $p_T^h$, so different notations
are reserved for them to avoid confusion.}. The quantity $F(x_g, k^g_T)$ denotes the unintegrated gluon distribution function (UGDF) 
describing interactions of the incoming quark with the target color field, which can be obtained by a Fourier transform of 
the dipole cross section $\sigma_{q\bar{q}}(\rho)$ (see Ref.~\cite{stastody} for more details). 

Integrating equation (\ref{eq:dy-dijet}) over the final hadron $h$ momentum, rapidity and relative angle between $G^*$ and $h$ 
one arrives at the inclusive gauge boson production cross section
\begin{eqnarray}
\frac{d \sigma(pp\rightarrow G^* X)}{dY d^2 p_T} & = & \int_{x_1}^1 \frac{dz}{z} 
\int d^2 k^g_T \sum_f x_p q_f(x_p,\mu_F)\, S_{\perp}\, F(x_g, k^g_T) \nonumber \\
& \times & \left\{  \frac{({\cal C}^G_f)^2  g_{v,f}^2}{2\pi} \left[  \left( 1 + (1-z)^2 \right) 
\frac{z^2 {k^g_T}^2}{\left[ p_T^2 + \epsilon_M^2  \right] 
\left[ ({\bf p}_T - z{\bf k}^g_T)^2 + \epsilon_M^2 \right]} \right.\right.\nonumber\\
& \quad & \qquad \qquad \left.\left. - z^2\epsilon_M^2 \left( \frac{1}{p_T^2 + \epsilon_M^2} - 
\frac{1}{({\bf p}_T - z{\bf k}^g_T)^2 + \epsilon_M^2} \right)^2 \right]\right.\nonumber\\
& \quad & \left. + \;  \frac{({\cal C}^G_f)^2  g_{a,f}^2}{2\pi} \left[  \left( 1 + (1-z)^2 \right) 
\frac{z^2 {k^g_T}^2}{\left[ p_T^2 + \epsilon_M^2  \right] 
\left[ ({\bf p}_T - z{\bf k}^g_T)^2 + \epsilon_M^2 \right]} \right.\right.\nonumber\\
& \quad & \qquad \qquad \left.\left.  - \frac{z^2 \epsilon_M^4}{M^2} 
\left( \frac{1}{p_T^2 + \epsilon_M^2} - \frac{1}{({\bf p}_T - z{\bf k}^g_T)^2 + 
\epsilon_M^2} \right)^2 \right] \right\} \,.
\label{eq:dy-boson}
\end{eqnarray}

Eqs.~(\ref{eq:dy-dijet}) and (\ref{eq:dy-boson}) allow us to construct the correlation function $C(\Delta \phi)$, 
which depends on the azimuthal angle difference $\Delta \phi$  between the trigger and associate particles. 
Experimentally, this coincidence probability is defined in terms of the yield of the correlated trigger and associated 
particle pairs $N_{pair} (\Delta \phi)$ and the trigger particle yield $N_{trig}$ as the following ratio: 
$C(\Delta \phi) = N_{pair} (\Delta \phi)/N_{trig}$. Therefore, azimuthal correlations are investigated 
through a coincidence probability defined in terms of a trigger particle, which could be either the gauge boson 
or the hadron. Here we assume the former as trigger particle, so that the correlation function is written as
\begin{eqnarray}
C(\Delta \phi) = \frac{ 2\pi\, \int_{p_T, p_T^h > p_T^{\rm cut}} dp_T p_T \; dp_T^h p_T^h \; 
\frac{d \sigma(pp \to h G^* X)}{d Y d y_h d^2p_T d^2p_T^h }} 
{\int_{p_T > p_T^{\rm cut}} dp_T p_T \; 
\frac{d \sigma(pp\rightarrow G^* X)}{dY d^2 p_T} }\,,
\label{corr}
\end{eqnarray}
where $p_T^{\rm cut}$ is the experimental low cut-off on transverse momenta of the resolved $G^*$ (or dilepton) 
and $h$, $\Delta \phi$ is the angle between them.

\subsection{Dipole cross section}

The main ingreadient of the dipole model is the dipole cross section $\sigma_{q\bar{q}}(\rho,x)$, which represents 
elastic scattering of a $q\bar q$ dipole of transverse separation $\rho$ at Bjorken $x$ 
off a nucleon \cite{zkl}. It is known to vanish quadratically $\sigma_{q\bar q}(\rho,x)\propto \rho^2$ as $\rho \to 0$ due to color screening 
which is the color transparency property \cite{zkl,BBGG,BM}. It cannot be predicted reliably from the first principles because of poorly known 
higher-order perturbative QCD corrections and non-perturbative effects. In particular, it should contain an information about non-linear QCD effects 
in the hadronic wave function (see e.g. Ref.~\cite{hdqcd}). In recent years several groups have constructed a number of viable 
phenomenological models based on saturation physics and fits to the HERA and RHIC data (see e.g. 
Refs.~\cite{GBW,iim,kkt,dhj,Goncalves:2006yt, buw, kmw, agbs, Soyez2007, bgbk, kt, ipsatnewfit, amirs}).

Since our goal is to extend previous DY studies to the kinematical range probed by the massive gauge boson production, 
where the main contribution comes from the small dipoles, in what follows we will consider two distinct phenomenological 
models which take into account the DGLAP evolution as well as the saturation effects. The first one is the model proposed 
in Ref.~\cite{bgbk}, denoted BGBK hereafter, where the dipole cross section is given by
\begin{equation}
 \sigma_{q\bar{q}} (\rho,x) = \sigma_0\left[1-\exp\left(- \frac{\pi^2}{\sigma_0 N_c} \rho^2 \alpha_s(\mu^2) xg(x, \mu^2)    \right)\, \right]\,\,,
\end{equation}
where $N_c=3$ is the number of colors, $\alpha_s(\mu^2)$ is the strong coupling constant at the scale $\mu^2$ which 
is related to the dipole size $\rho$ as $\mu^2=C/\rho^2 + \mu_0^2$, with $C$, $\mu_0$ and $\sigma_0$ parameters 
fitted to HERA data. Moreover, in this model, the gluon density evolves according to DGLAP equation \cite{dglap} accounting for gluons only
\begin{equation}
\frac{\partial xg(x,\mu^2)}{\partial \ln \mu^2 } = \frac{\alpha_s(\mu^2)}{2\pi} \int_x^1 \dd z  P_{gg}(z) \frac{x}{z} g(\frac{x}{z}, \mu^ 2)\,, \label{dglap}
\end{equation}
where the gluon density at initial scale $\mu_0^2$ is parametrized as
\begin{equation}
xg(x,\mu_0^2) = A_g x^{-\lambda_g} (1-x)^{5.6}\,.
\end{equation}
The best fit values of the model parameters are the following: $A_g = 1.2$, $\lambda_g = 0.28$, $\mu_0^2 = 0.52$ GeV$^{2}$, $C = 0.26$ and 
$\sigma_0 = 23$ mb. This model was generalized in Ref.~\cite{kmw} in order to take into account the impact parameter dependence of the dipole 
cross section and to describe exclusive observables at HERA. In this model, denoted as IP-SAT hereafter, the corresponding dipole cross section is given by
\begin{eqnarray}
\sigma_{q\bar{q}} (\rho,x)  = 2\,\int d^2b\, \left[1-\exp\left(- \frac{\pi^2}{2 N_c} \rho^2 \alpha_s(\mu^2) xg(x, \mu^2) T_G({\bf b})   \right)\, \right]
\end{eqnarray} 
with the evolution of the gluon distribution given by Eq.~(\ref{dglap}). The Gaussian impact parameter dependence is given by 
$T_G({\bf b})=(1/2\pi B_G) \exp(-b^2/2 B_G)$, where $B_G$ is a free parameter extracted from the $t$-dependence of 
the exclusive $ep$ data. The parameters of this model were updated in Ref.~\cite{ipsatnewfit} by fitting to the recent 
high precision HERA data \cite{heradata} providing the following values: $A_g = 2.373$, $\lambda_g = 0.052$, $\mu_0^2 = 1.428$ GeV$^{2}$, 
$B_G = 4.0$ GeV$^{2}$ and $C = 4.0$. 
\begin{table}[!h]
\footnotesize 
\begin{center}
\begin{tabular}{|c@{\quad}||c@{\quad}|c@{\quad}|c@{\quad}|c@{\quad}|}
\hline

 \bf{$\sqrt{s}$ (TeV)} & {\bf GBW}  & {\bf BGBK}  & {\bf IP-SAT}   & {\bf DATA  (nb)}  \\ [0.5ex] \hline \hline
     7           & 0.950  &  1.208     &  0.986    & \parbox[t]{5cm}{$0.937 \pm 0.037$ (ATLAS) \\ $0.974 \pm 0.044$ (CMS) } \\ \hline
  
 
     8           & 1.083  &  1.427   &  1.183     & $1.15 \pm 0.37$ (CMS) \\ \hline

     14          & 1.852  &  2.797   &   2.514  & -- \\ \hline \hline

\end{tabular}
\end{center}
\caption{Comparison between the GBW, BGBK and IP-SAT predictions for the total cross sections for $Z^0$ boson 
production at different values of the c.m. energy. The experimental results are from Refs.~\cite{cms_data,atlas_data,cms_data_8tev}.  
The cross sections are given in nanobarns. }
\label{tab-1}
\end{table}

Finally, for comparison with previous results available in the literature, we also consider the Golec-Biernat--Wusthoff (GBW) model \cite{GBW} 
based upon a simplified saturated form
\begin{equation}
\label{gbw}
\sigma_{q\bar{q}}(\rho,x) = \sigma_0\left(1 - e^{-\frac{\rho^2 Q_s^2(x)}{4} } \right)\,
\end{equation}
with the saturation scale
\begin{equation}
Q_s^2(x) = Q_0^2\left( \frac{x_0}{x} \right)^\lambda \,,
\label{satsca}
\end{equation}
where the model parameters $Q_0^2 = 1$ GeV${}^2$, $x_0 = 3.04\times10^{-4} \, (4.01\times10^{-5})$, $\lambda = 0.288 \, (0.277)$ and 
$\sigma_0 = 23.03 \, (29)$ mb were obtained from the fit to the DIS data without (with) the contribution of the charm quark, respectively.
\begin{figure}[!h]
\large
\begin{center}
\scalebox{0.72}{\includegraphics{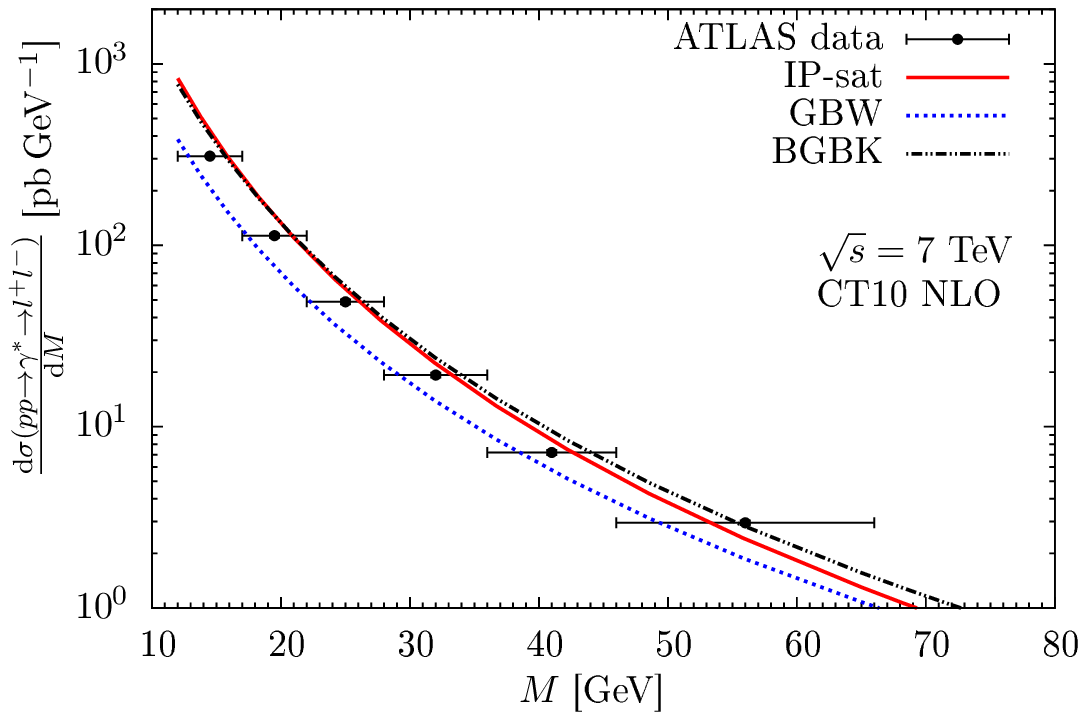}}
\scalebox{0.72}{\includegraphics{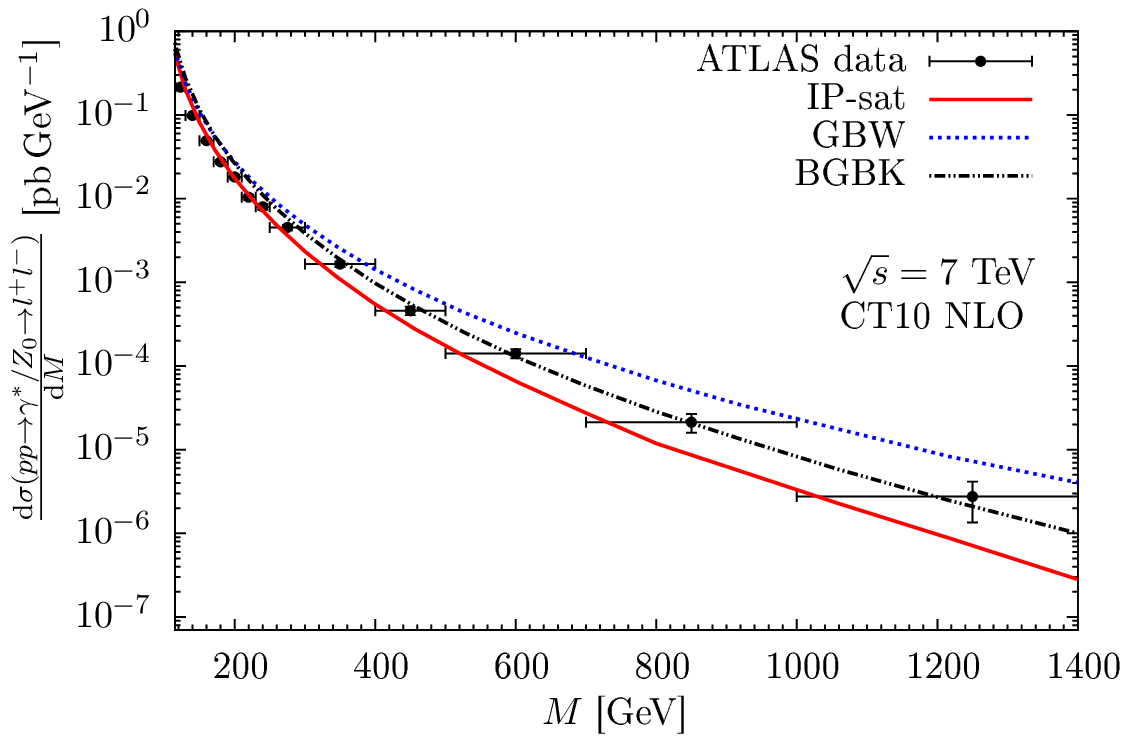}}
\caption{The DY pair invariant mass distribution of the process 
$pp\rightarrow \gamma^*/Z^0 \rightarrow l \bar l$ at $\sqrt{s}=7$ TeV 
in low (left panel) and high (right panel) invariant mass ranges compared 
to the data from the ATLAS Collaboration \cite{ATLASlowmass,ATLAShighmass} 
for three different parametrisations of the dipole cross section.}
\label{fig:DYmass}
\end{center}
\end{figure}
\normalsize
\begin{figure}[!h]
\large
\begin{center}
\scalebox{0.72}{\includegraphics{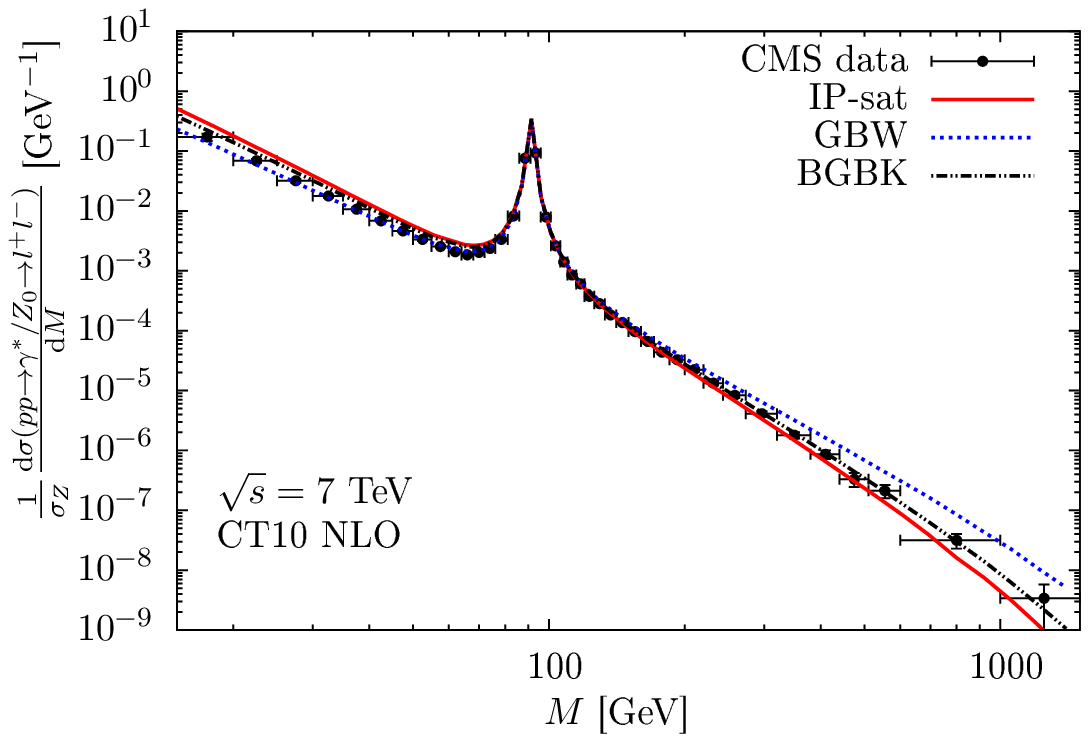}}
\scalebox{0.72}{\includegraphics{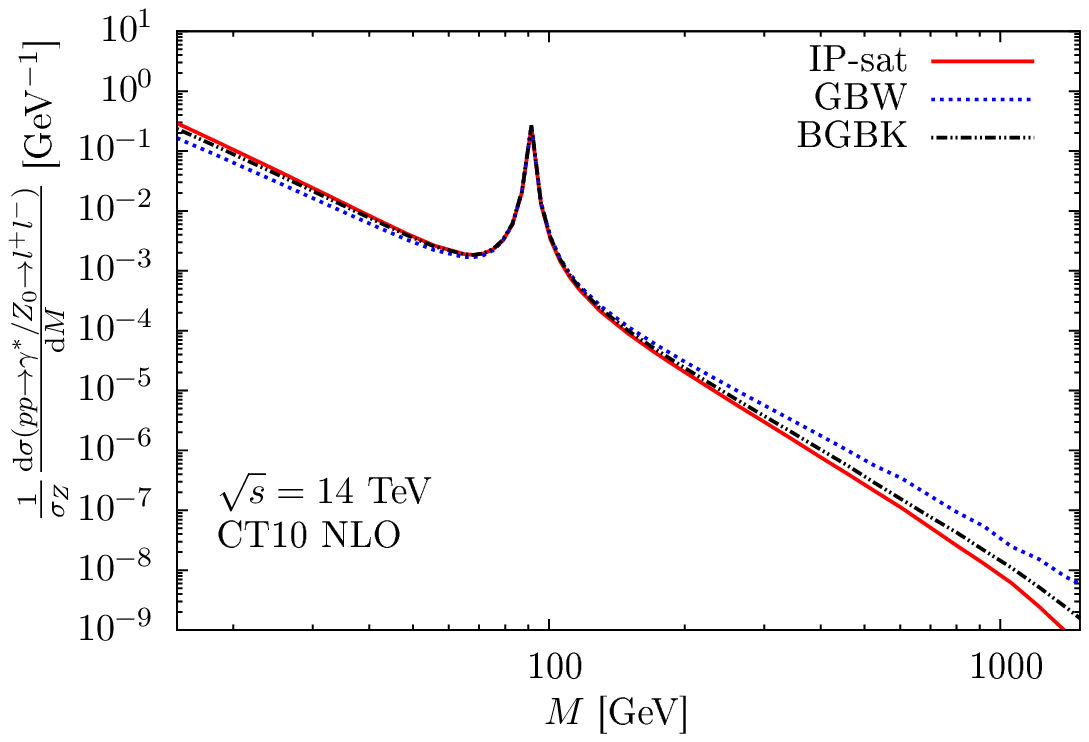}}
\caption{The DY pair invariant mass distribution of the process 
$pp\rightarrow \gamma^*/Z^0 \rightarrow l \bar l$ at $\sqrt{s} = $ 7 TeV 
compared to the data from the CMS collaboration \cite{cmsDYmass} 
for three different parametrisations of the dipole cross section in the left panel. 
The corresponding predictions are shown for $\sqrt{s} = $ 14 TeV in the right panel. }
\label{fig:Zmass}
\end{center}
\end{figure}
\normalsize
\begin{figure}[!h]
\large
\begin{center}
\scalebox{0.65}{\includegraphics{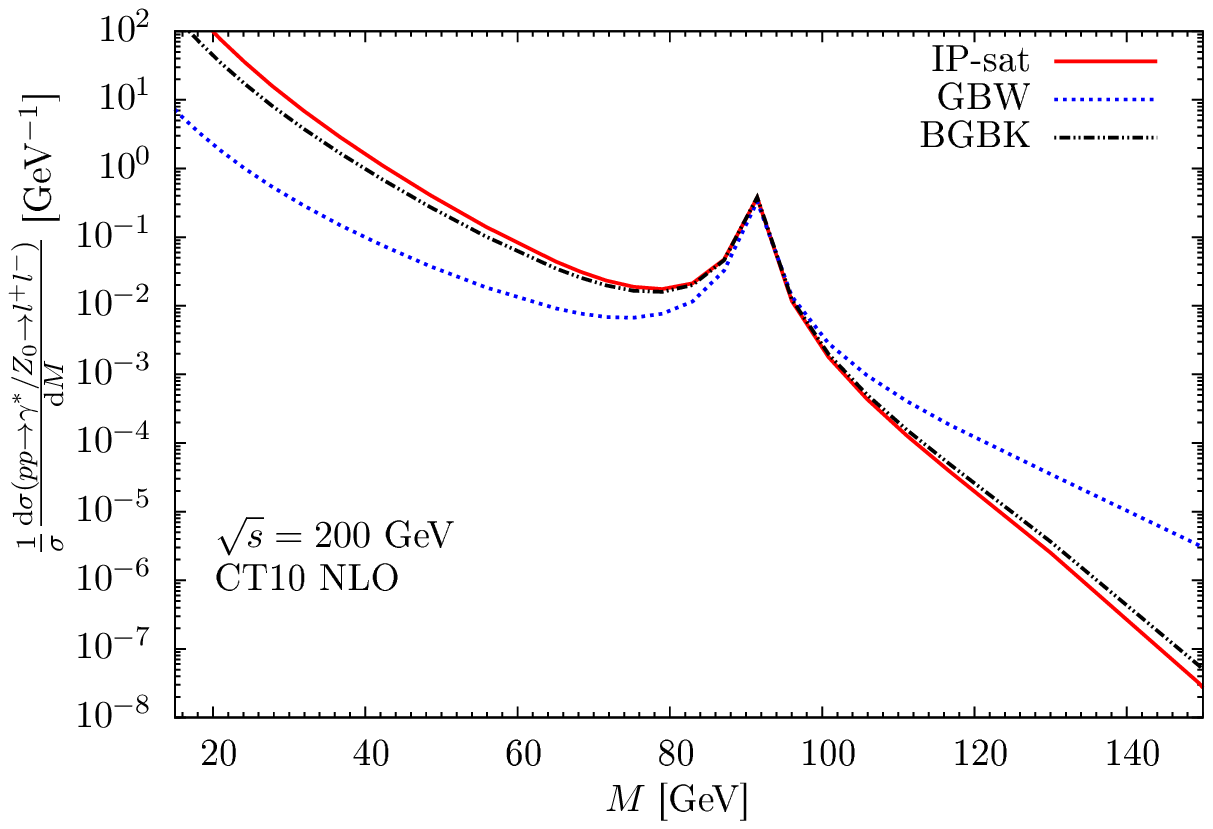}}
\scalebox{0.65}{\includegraphics{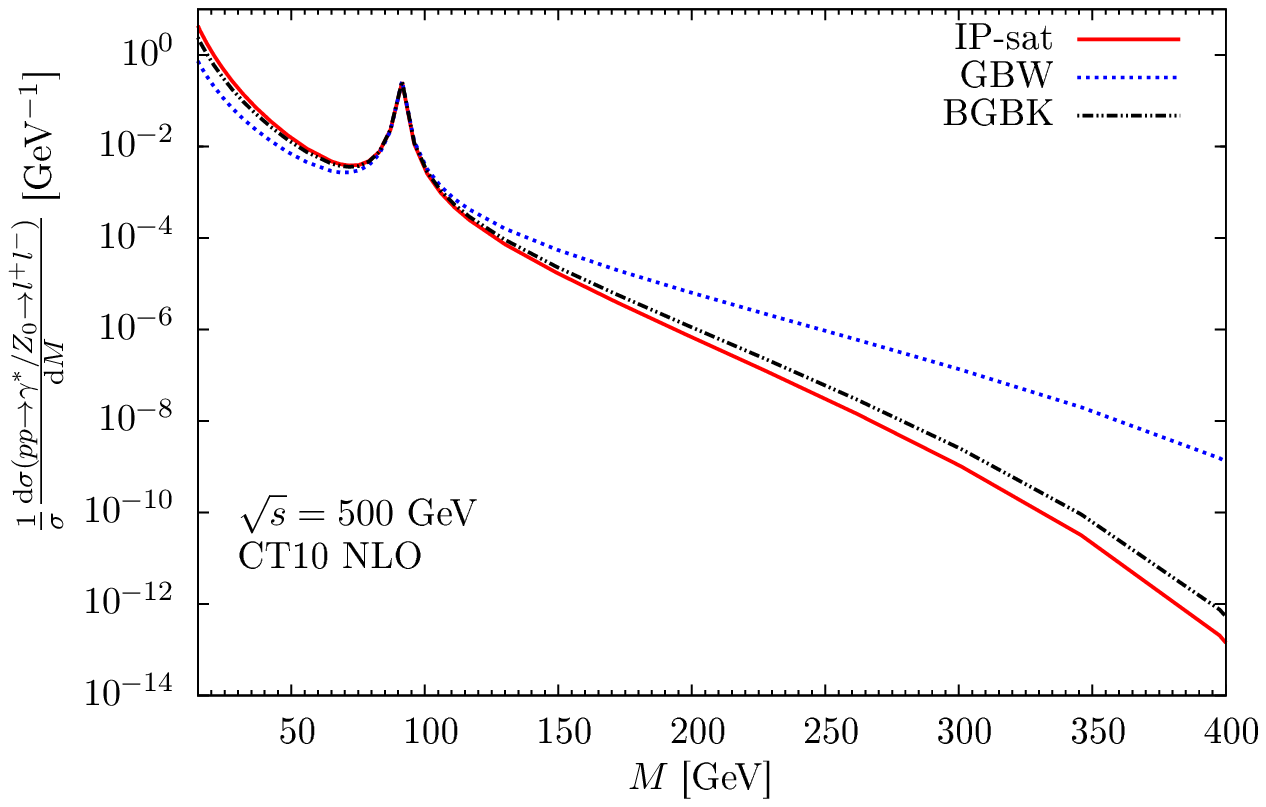}}
\caption{The DY pair invariant mass distribution of the process 
$pp\rightarrow \gamma^*/Z^0 \rightarrow l \bar l$ at 
RHIC Run I ($\sqrt{s} = 200$ GeV) and II ($\sqrt{s} = 500$ GeV) 
energies for three different parametrisations of the dipole 
cross section.}
\label{fig:Zmass_rhic}
\end{center}
\end{figure}
\normalsize

%
%
%
%
%
\section{Numerical results}
\label{res}
%
%
%
%
%

In what follows, we present our predictions for the DY pair production in the process $pp\rightarrow \gamma^*/Z^0 \rightarrow l \bar l$ 
obtained by using the color dipole formalism and the three phenomenological models for the dipole cross section discussed in the previous Section. 
Following Ref.~\cite{GBW}, in calculations of the DY pair production cross sections we take the quark mass values to be $m_u = m_d = m_s = 0.14$ GeV, 
$m_c = 1.4$ GeV and $m_b = 4.5$ GeV, and employ the CT10 NLO parametrization for the projectile quark PDFs \cite{ct10} with the factorization 
scale $\mu_F=M$.

%
%
%
\subsection{Predictions for DY pair production cross sections}
%
%
%

To start with, in Table~\ref{tab-1} we present our results for the total $Z^0$ production cross sections for several parameterisations of the dipole cross section 
and different c.m. energies accessible at the LHC. The GBW model gives the cross section value at $\sqrt{s}=7$ TeV smaller than that obtained by using 
the IP-SAT model which correctly treats the region of large transverse momenta due to DGLAP evolution (small dipoles). The DY cross section obtained by using 
the BGBK model turns out to be somewhat higher than the 7 TeV data while the GBW and IP-SAT are well within the error bars while all three models describe 
$\sqrt{s}=8$ TeV data rather well. It is worth to mention that in comparison to Ref.~\cite{Basso} we obtain somewhat larger $Z^0$ cross sections for the GBW 
case due to an extra factor $x_1/(x_1+x_2)$ and the use of the NLO quark PDFs.

Such a fairly good description of the LHC data on the total $Z^0$ cross section naturally motivates a more detailed analysis of the DY cross section differential in 
dipleton invariant mass, rapidity and transverse momentum. In Fig.~\ref{fig:DYmass} we compare our predictions for the invariant mass distributions 
with the recent ATLAS data in low and high $M$ ranges. We conclude that the dipole cross section parameterisations including the DGLAP evolution 
via the gluon PDF describe the DY data better compared to the GBW model, especially at high invariant masses. The large invariant mass region prefers 
a fairly large $Z^0$ contribution as well as its interference with $\gamma^*$ in the considered dilepton channel. This is clearly seen in the left panel of 
Fig.~\ref{fig:Zmass} where we present a comparison of the CMS data with predictions using three different parameterisations of the dipole cross section.
Indeed, the DGLAP evolution, included in both IP-SAT and BGBK models leads to a better agreement with the data at large $M$ in comparison with the GBW 
model. Our predictions for $pp$ collisions at $\sqrt{s} = $14 TeV are presented in the right panel of Fig.~\ref{fig:Zmass}. Considering that similar measurements
can be performed at RHIC, in Fig.~\ref{fig:Zmass_rhic} we show our results for the dilepton invariant mass distributions corresponding to the c.m. energies 
of Run I ($\sqrt{s} = 200$ GeV) and II ($\sqrt{s} = 500$ GeV). While the $M$-distributions corresponding to the IP-SAT and BGBK models are rather 
close to each other, the GBW predictions significanty differ from them away from the $Z^0$ peak, in both low and, especially, 
high invariant mass ranges.
\begin{figure}[t]
\large
\begin{center}
\scalebox{0.72}{\includegraphics{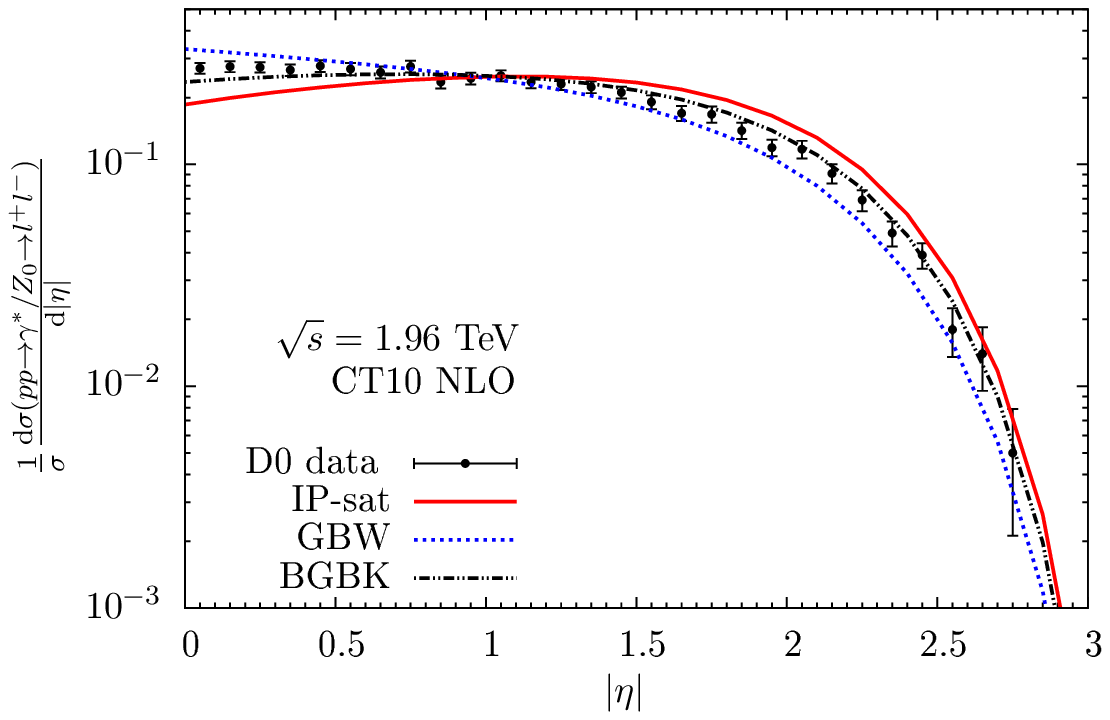}}
\scalebox{0.72}{\includegraphics{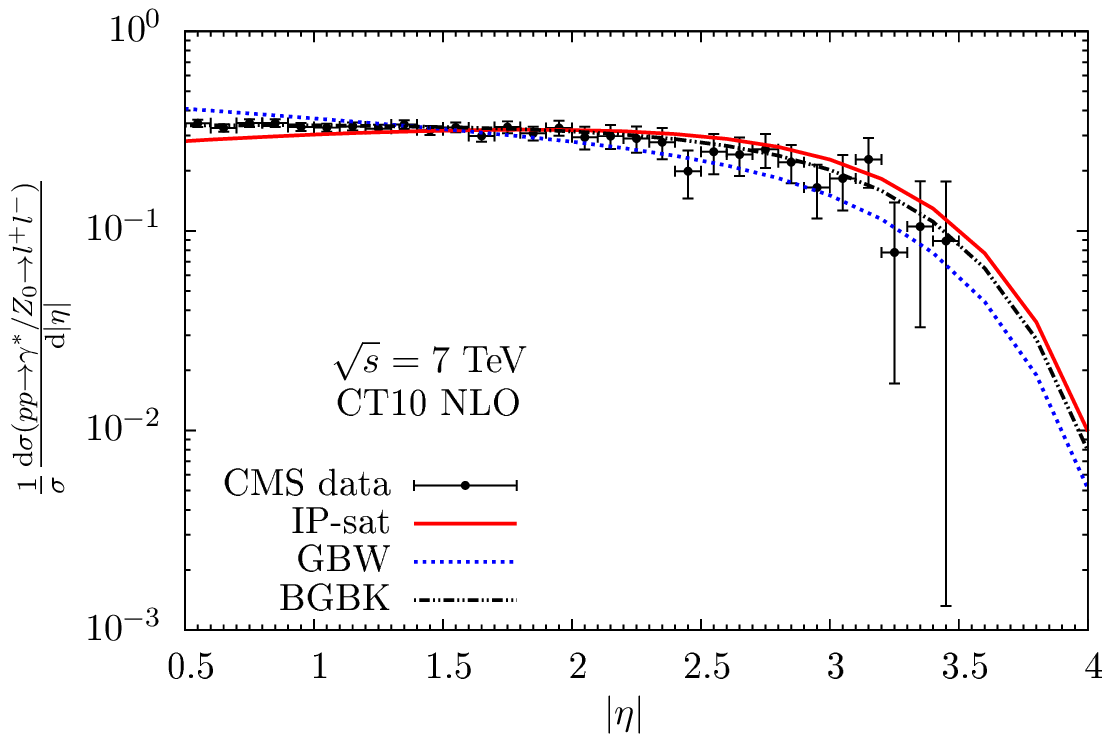}}
\scalebox{0.72}{\includegraphics{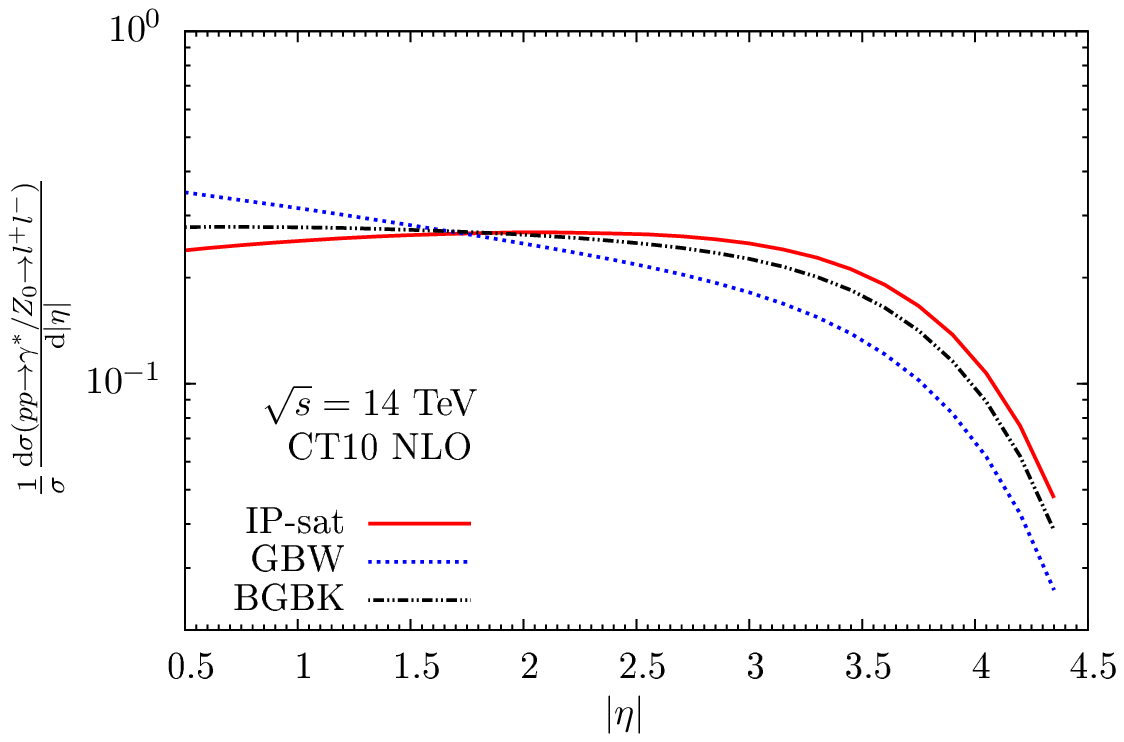}}
\caption{The $Z^0$ boson rapidity distribution for different center of mass energies: 
$\sqrt{s} = $ 1.96 TeV (top left panel), 7 TeV (top right panel) and 14 TeV (bottom panel) 
versus data from the D0 \cite{d0-07} and CMS \cite{Zcms} Collaborations. }
\label{fig:Zeta}
\end{center}
\end{figure}
\normalsize

In Fig.~\ref{fig:Zeta} we present our predictions for the $Z^0$ boson rapidity distribution at different c.m. energies corresponding 
to Tevatron $\sqrt{s} = $ 1.96 TeV (top left panel) and LHC Run I $\sqrt{s} = $ 7 TeV (top right panel). These results show that the 
IP-SAT and GBW models deviate from data in the central rapidity region while the BGBK predictions come closer to the data in the 
whole rapidity range. It is worth to emphasize that specifically for the Tevatron energy and for central rapidities, the results
are rather sensitive to the behaviour of the dipole cross section at large values of $x$, which is not under control in the considered 
formalism. At larger rapidities we obtain a reasonable description of the data though. The future LHC data at $\sqrt{s} = $14 TeV 
at large $M>m_Z$ can be used to put even stronger constraints on the dipole model parametrisations (see the bottom panel 
in Fig.~\ref{fig:Zeta}) whose predictions significantly differ in the large rapidity region.
\begin{figure}[t]
\large
\begin{center}
\scalebox{0.72}{\includegraphics{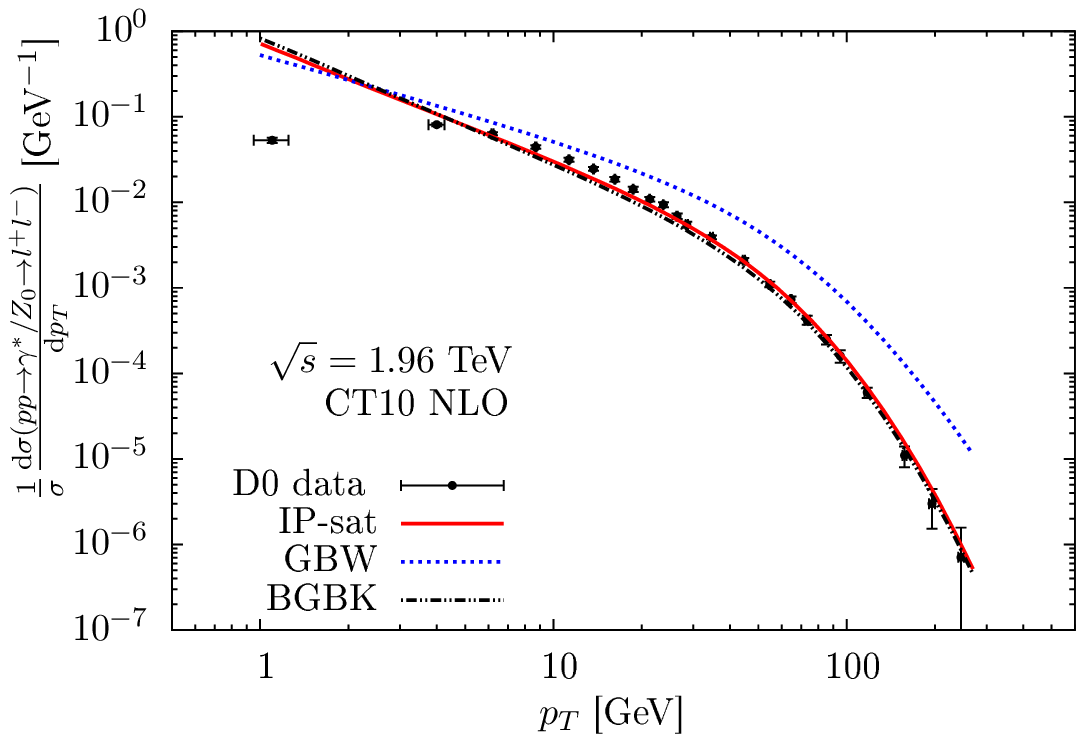}}
\scalebox{0.72}{\includegraphics{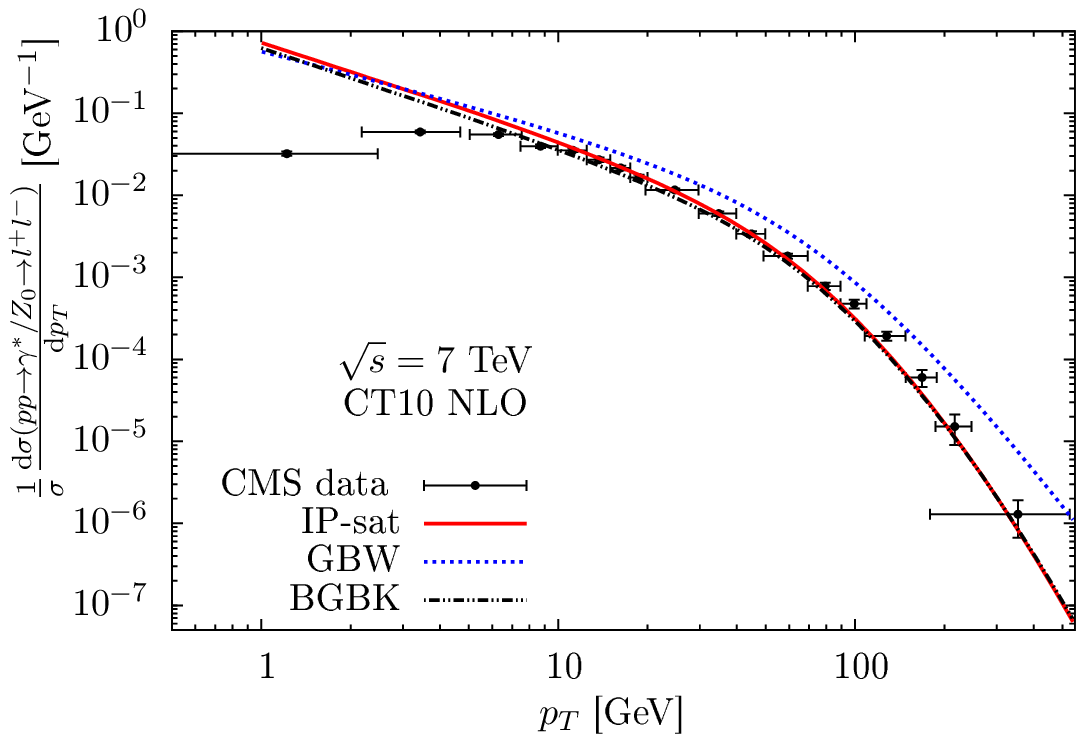}}
\scalebox{0.72}{\includegraphics{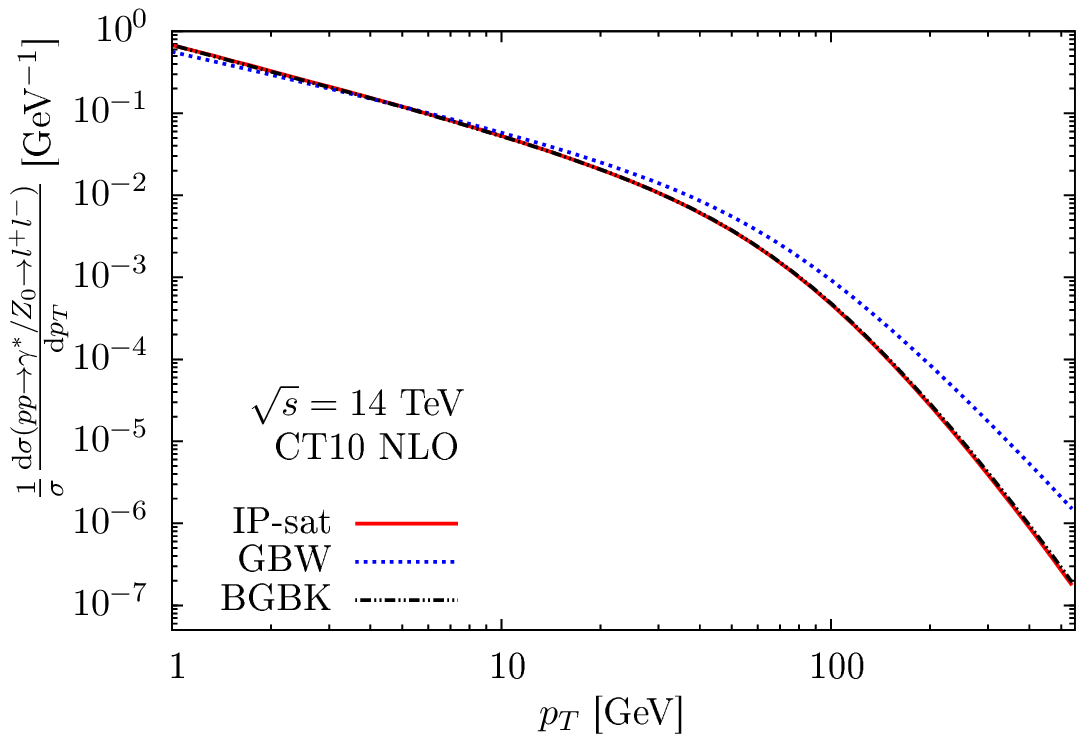}}
\caption{The transverse momentum distributions of $Z^0$ bosons in $pp$ collisions at 
$\sqrt{s} = $ 1.96 TeV (top left panel), 7 TeV (top right panel) and 14 TeV (bottom panel) versus
data from the D0 \cite{d0-08} and  CMS  \cite{Zcms} Collaborations.}
\label{fig:Zpt}
\end{center}
\end{figure}
\normalsize

We turn now to a discussion of the transverse momentum distributions of the DY pair production cross section. 
In what follows, we take into account that heavy gauge boson and highly virtual photon production implies 
that typical dipole separations are small, i.e. $\rho  \sim 1/\tau\ll 1/Q_s(x_2)$ ($\alpha\ll 1$). In this case, we can take
the quadratic form of the dipole cross section as follows
\begin{eqnarray}
\sigma(\rho,x) \approx \omega(x) \rho^2\,, \qquad \omega(x) = \frac{\pi^2}{2 N_c}\, \alpha_s(\mu^2)\, xg(x, \mu^2)\, T_G({\bf b}) \,,
\end{eqnarray}
for the IP-SAT model (a similar analysis can be done for the GBW and BGBK models), such that
the Fourier integral in Eq.~(\ref{ptdistcc}) can be performed analytically.
Then the square bracket in Eq.~(\ref{ptdistcc}) can be written as
\begin{equation}
\sigma_{q\bar{q}}(\alpha {\bm\rho}_1,x ) + \sigma_{q\bar{q}}(\alpha {\bm\rho}_2,x ) - 
\sigma_{q\bar{q}}(\alpha |{\bm\rho}_1- {\bm\rho}_2|,x) \approx \alpha^2 ({\bm\rho}_1 \cdot {\bm\rho}_2)\omega(x)\,.
\end{equation}
The $\rho$-dependent parts of the gauge boson wave functions in Eq.~(\ref{VV}) lead to
the following two Fourier integrals in the DY cross section (for more details, see Appendix B 
in Ref.~\cite{dynuc})
\begin{align}
J_1(p_T, \tau) = & \int d^2\rho_1 d^2\rho_2\; ({\bm\rho}_1 \cdot {\bm\rho}_2)\;
{\rm K}_0\left(\tau \rho_1\right)\,{\rm K}_0\left(\tau \rho_2\right)
\exp[i{\bf p}_T \cdot ({\bm\rho}_1 - {\bm\rho}_2)] = 16 \pi^2 \frac{p_T^2}{(\tau^2 + p_T^2)^4}\\
J_2(p_T, \tau) = & \int d^2\rho_1 d^2\rho_2\; \frac{({\bm\rho}_1 \cdot {\bm\rho}_2)^2}{\rho_1 \rho_2} \;
{\rm K}_1\left(\tau \rho_1\right)\,{\rm K}_1\left(\tau \rho_2\right)
\exp[i{\bf p}_T \cdot ({\bm\rho}_1 - {\bm\rho}_2)] = 8 \pi^2 \frac{\tau^4 + p_T^4}{\tau^2(\tau^2 + p_T^2)^4}\,.
\end{align} 

The considered small dipole limit is valid as long as the hard scale $\tau\sim \mu_F\sim M$ is large enough compared 
to the saturation scale $Q_s$. One should note, at low transverse momenta (e.g. $p_T\lesssim 3-5$ GeV at the LHC) 
the contribution of an intrinsic primordial transverse momentum of the projectile quark in the incoming proton wave function 
and the corresponding Sudakov suppression can be important \cite{Sudakov}. We postpone the analysis of these effects within 
the dipole formalism for a future investigation.

\begin{figure}[t]
\large
\begin{center}
\scalebox{0.72}{\includegraphics{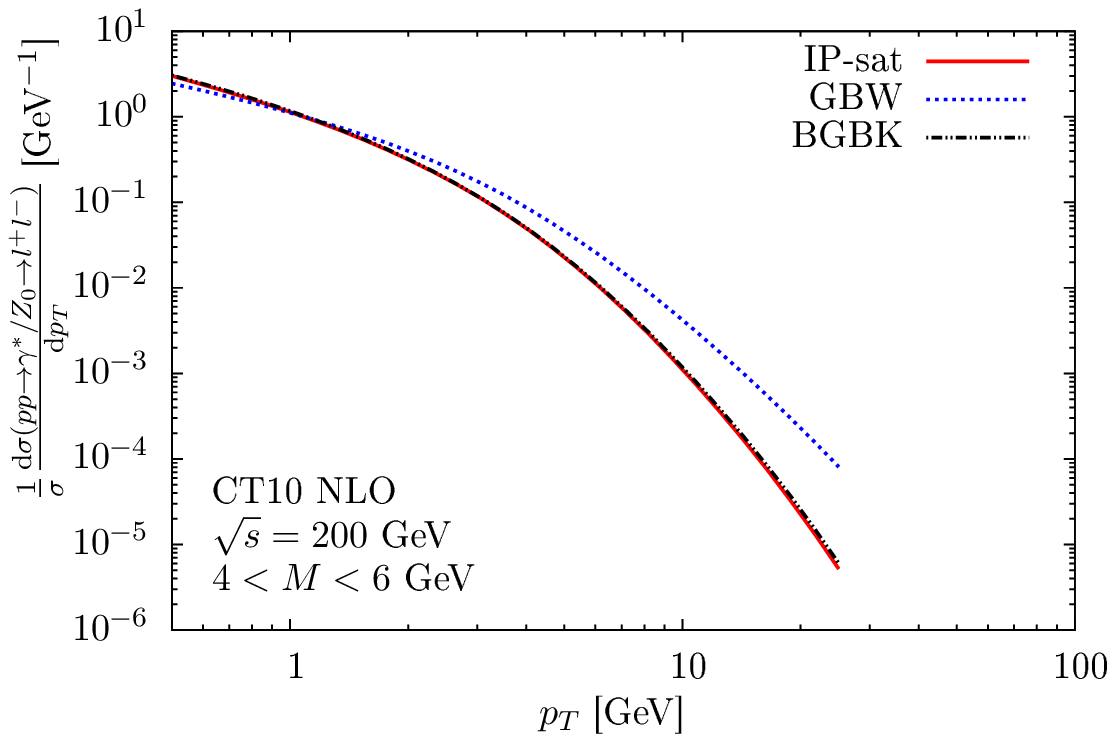}}
\scalebox{0.72}{\includegraphics{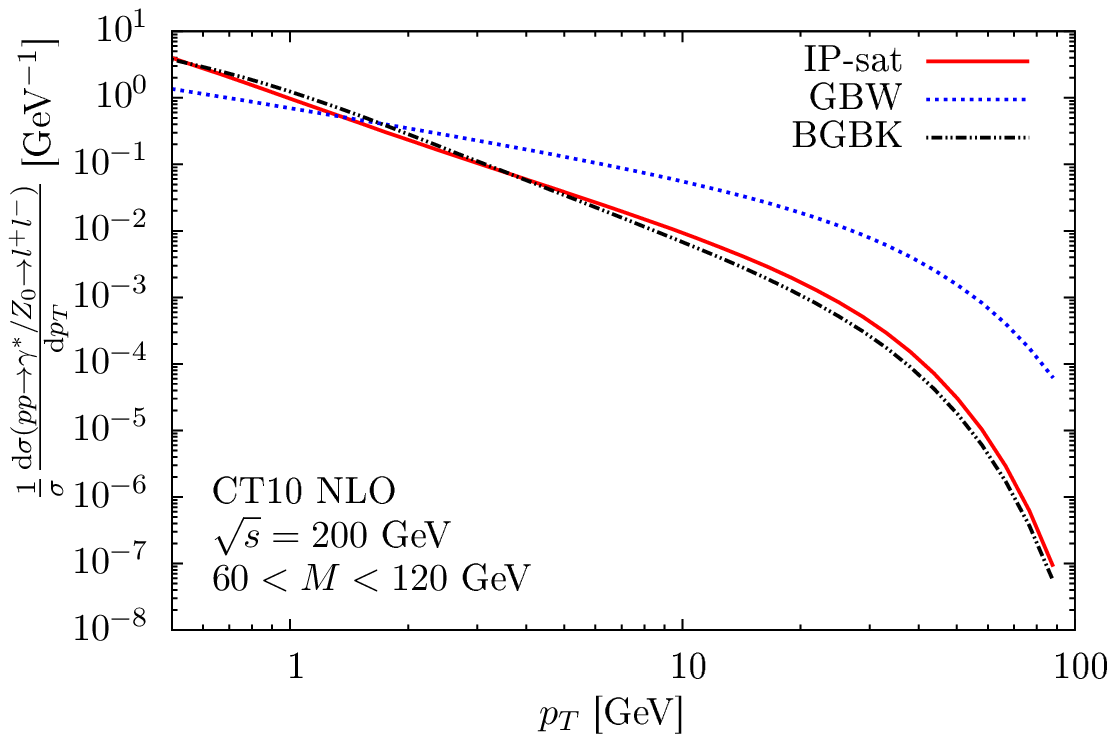}}\\
\scalebox{0.72}{\includegraphics{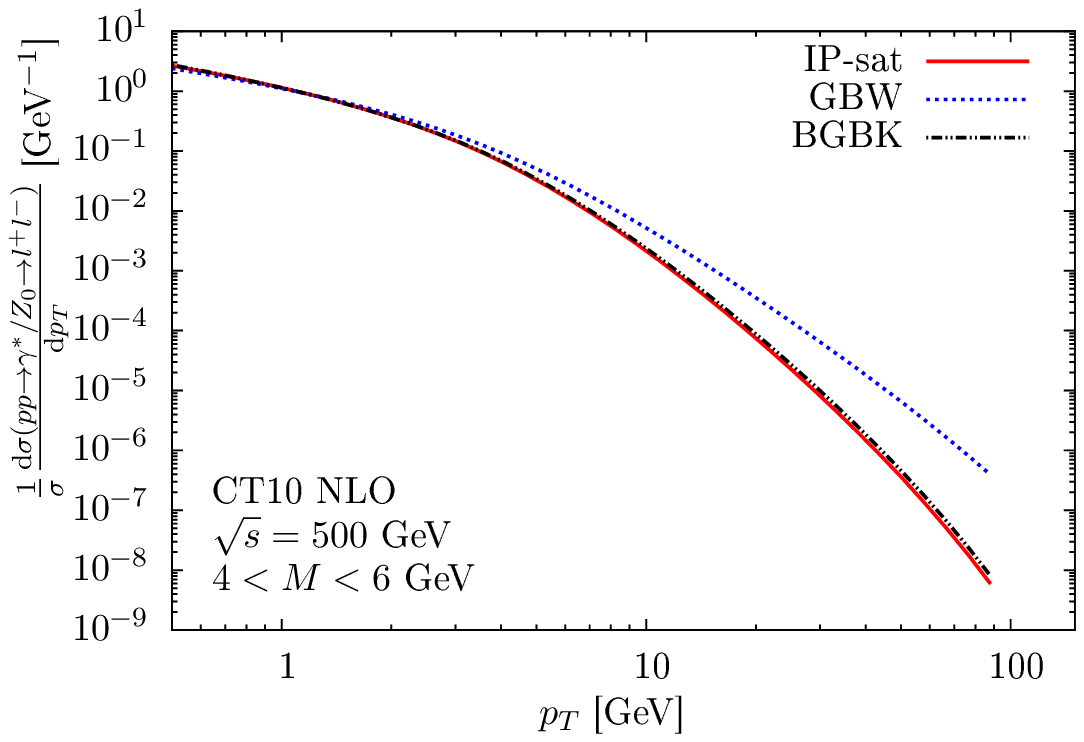}}
\scalebox{0.72}{\includegraphics{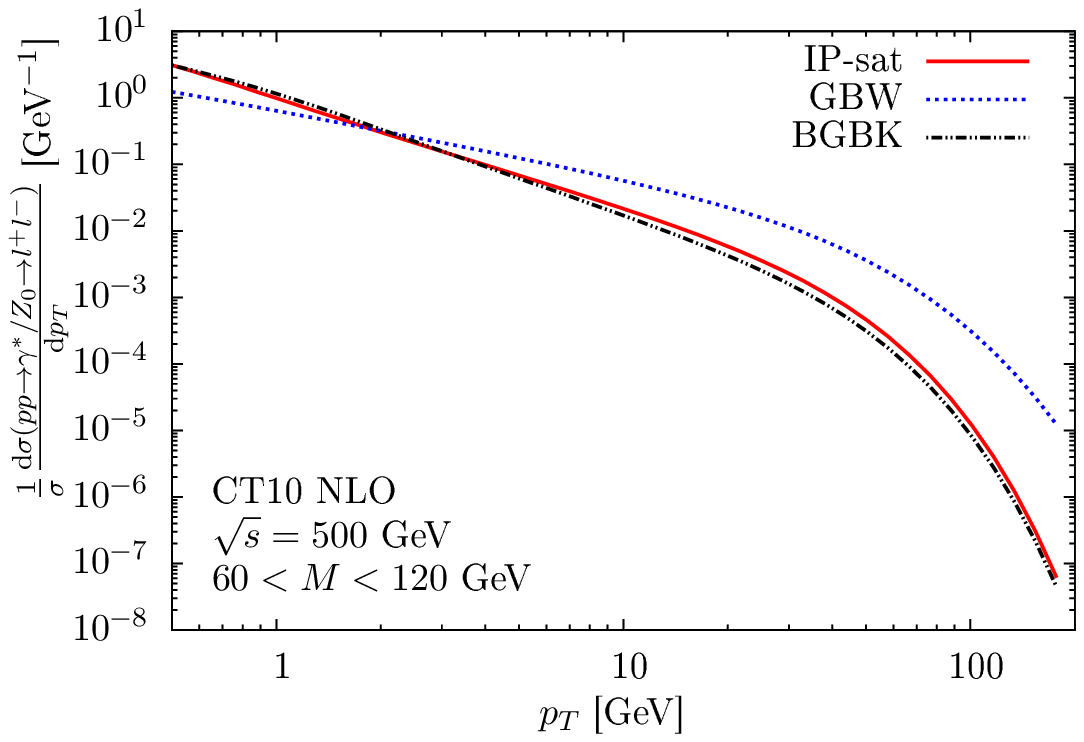}}
\caption{The transverse momentum distributions for DY production 
in $pp$ collisions at RHIC ($\sqrt{s}= 200, 500$ TeV) considering 
two different invariant mass ranges.}
\label{fig:Zpt_rhic}
\end{center}
\end{figure}

We present our predictions for the dilepton $p_T$ distribution in Fig.~\ref{fig:Zpt} for $pp$ collisions at various c.m. energies: 
$\sqrt{s} = $ 1.96 TeV (top left panel), 7 TeV (top right panel) and 14 TeV (bottom panel). As was anticipated above, 
these predictions do not describe the experimental data in the low $p_T$ region. On the other hand, at large $p_T>5$ GeV 
the data are well described by the DGLAP-evolved dipole models IP-SAT and BGBK, but not GBW. In Fig.~\ref{fig:Zpt_rhic} we present 
the corresponding distributions at RHIC energies considering two different ranges of invariant masses. At low invariant masses, 
where the photon contribution strongly dominates, the IP-SAT and GBW models give very similar predictions. The difference between 
these models rises with $p_T$ and $M$. At large invariant masses probing the $Z^0$ peak region, the predictions differ even more 
significantly, and such a tendency continues to higher $M$. In this kinematical range, the DGLAP evolution makes the predictions 
more reliable at both low (RHIC) and high (LHC) energies.
\normalsize

%
%
%
\subsection{Predictions for the azimuthal correlation function in $pp$ collisions}
%
%
%

Finally, let us consider the correlation function $C(\Delta \phi)$ defined 
by Eq.~(\ref{corr}). This observable has been studied in Ref.~\cite{stasto} 
for the DY+pion production in proton(deuteron)-nucleus collisions 
at RHIC and LHC energies taking into account saturation effects and 
considering only the virtual photon contribution to dilepton production,
$\gamma^*\to l\bar l$. The authors have demonstrated that at variance 
to the near-side peak ($\Delta \phi = 0$) distribution, 
which is dominated by the leading jet fragmentation, 
the away-side peak ($\Delta \phi = \pi$) follows from back-to-back jets 
produced in the hard 2 $\to $ 2 scattering. Moreover, 
since low-$x$ gluons in the target dominate and carry a typically 
large transverse momentum of the order of the saturation scale, 
the transverse momentum imbalance of back-to-back jets increases at high energies.
So saturated gluons from the target tend to smear the back-to-back structure 
and suppress the away-side peak in the $\Delta \phi$ distribution. 

There are two main important results coming from the analysis in Ref.~\cite{stasto}. 
The first one is the prediction of a double peak in the correlation function 
distribution around $\Delta \phi = \pi$ with a dip at $\Delta \phi = \pi$. 
The second shows that such a behaviour of $C(\Delta\phi)$ is not 
strongly dependent on the large transverse momentum tail 
of the UGDF, which was used in calculations. 
The latter conclusion implies that it is possible 
to get realistic predictions employing also the GBW model 
for the dipole cross section. 
In this case, numerical calculations are significantly simplified 
and the UGDF has the following analytical form,
%
\begin{equation}
F(x_g,k^g_T) = \frac{1}{\pi Q_s^2(x_g) }\, e^{-{k^g_T}^2/Q_s^2(x_g)} \,,
\end{equation} 
%
with the saturation scale given by Eq.~(\ref{satsca}). 
In what follows, we study the correlation function $C(\Delta \phi)$ 
assuming the GBW model for the UGDF, 
the CT10 NLO parametrization for parton distributions 
and the Kniehl-Kramer-Potter (KKP) fragmentation function 
$D_{h/f}(z_h,\mu_F^2)$ of a quark with a flavor $f$ into a neutral pion $h=\pi^0$ \cite{kkp}. 
Moreover, we assume that a minimal transverse momentum ($p_T^{\rm cut}$) for the gauge boson $G$ and pion in 
Eq.~(\ref{corr}) is the same and is equaled 
to 1.5 and 3.0 GeV for RHIC and LHC energies, respectively. 
\begin{figure}[!h]
\large
\begin{center}
\scalebox{0.7}{\includegraphics{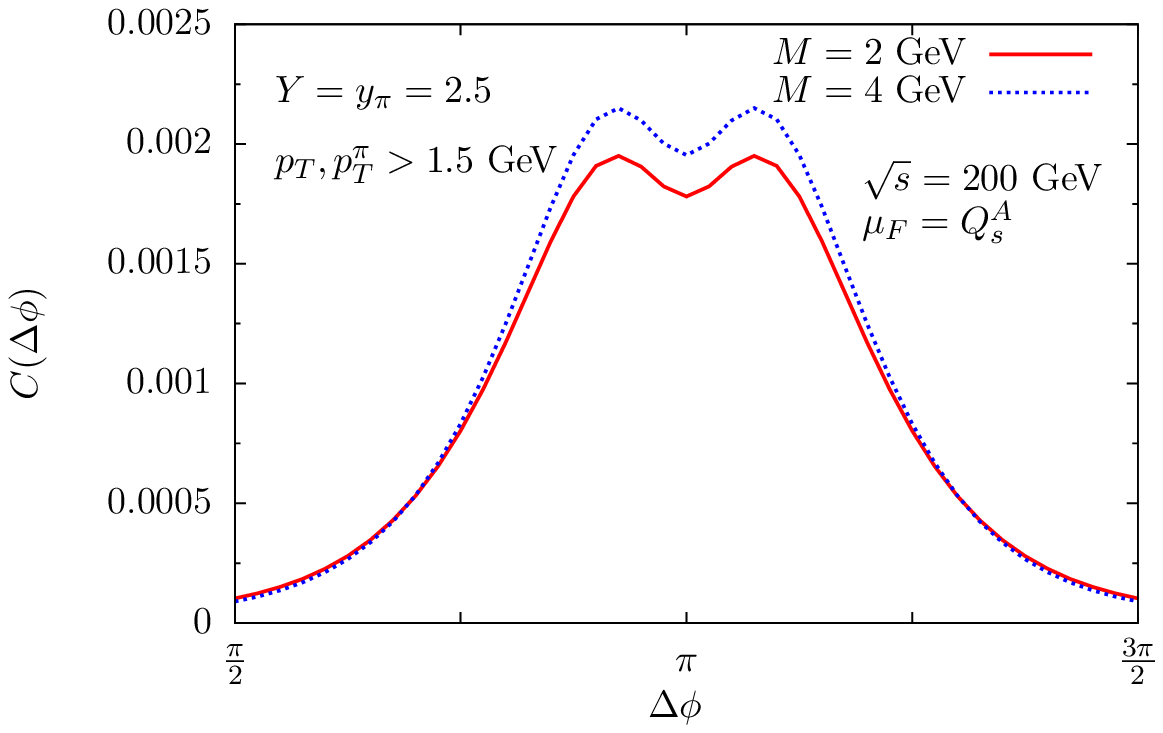}}
\scalebox{0.7}{\includegraphics{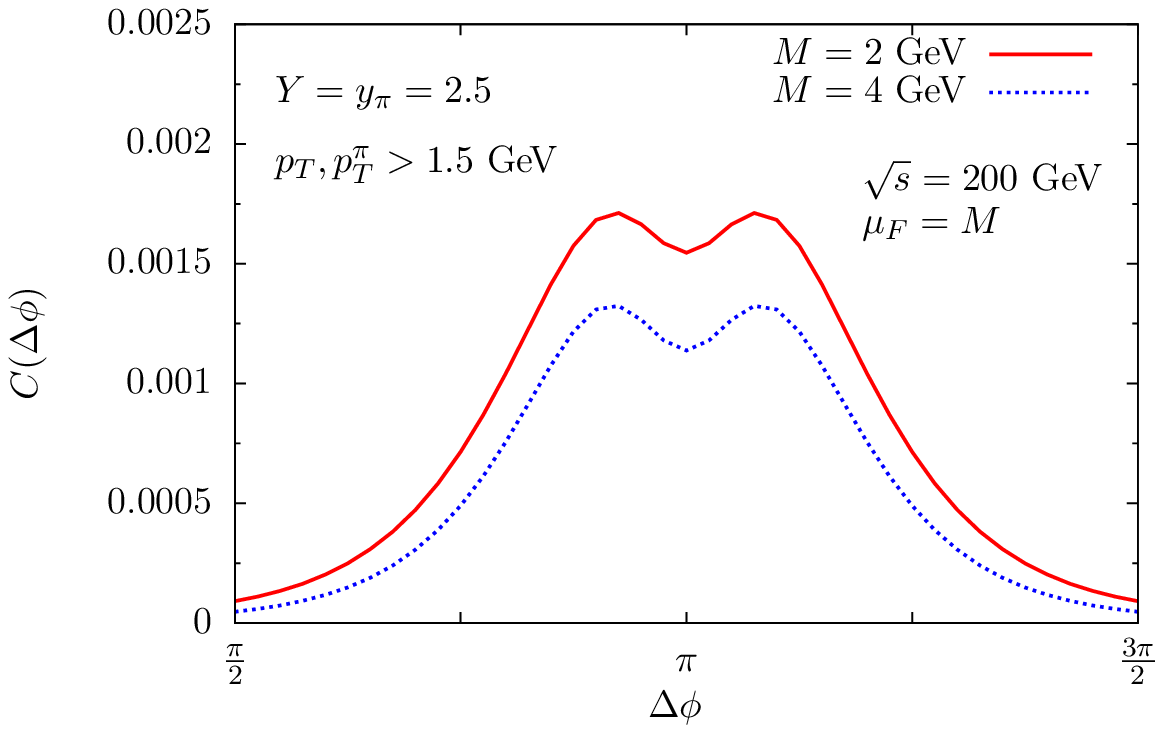}}
\caption{The correlation function $C(\Delta \phi)$ for the associated DY pair and pion production 
in $dAu$ collisions at RHIC ($\sqrt{s}=200$ GeV) assuming that factorization scale is given by 
the nuclear saturation scale $\mu_F=Q_s^A$ (left panel) or by the dilepton invariant mass $\mu_F=M$ (right panel). }
\label{fig:CP_rhic200}
\end{center}
\end{figure}
\normalsize
\begin{figure}[!h]
\large
\begin{center}
\scalebox{0.7}{\includegraphics{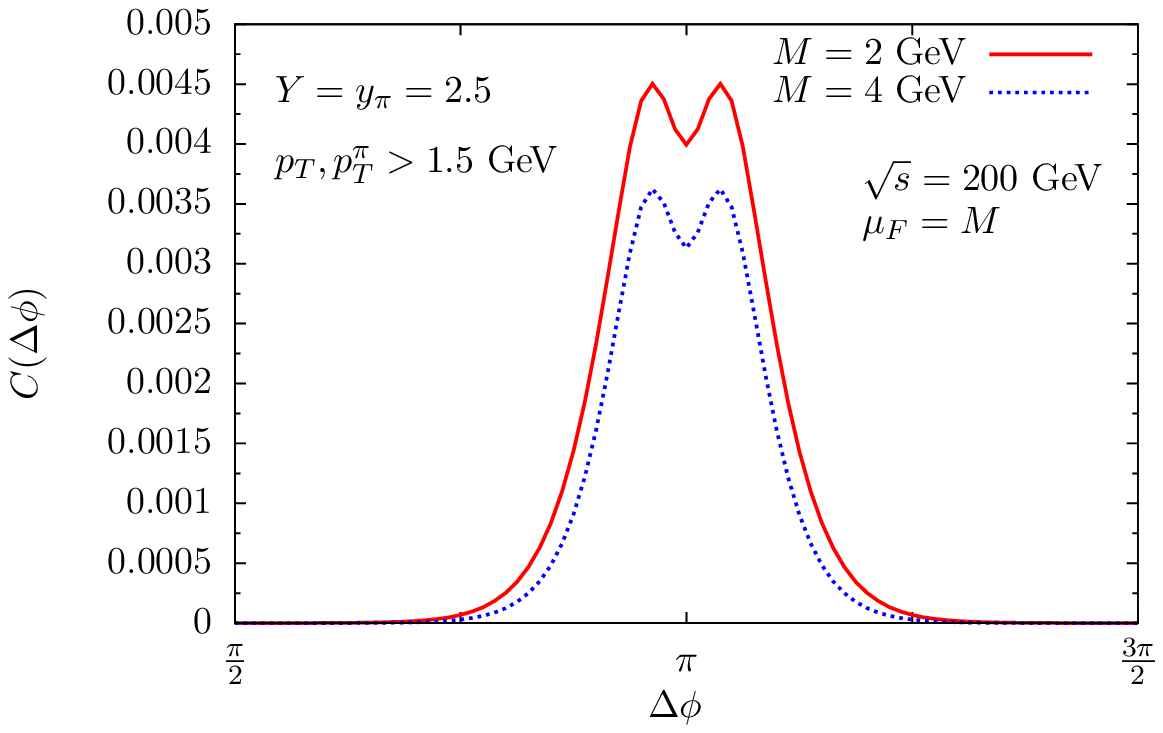}}
\scalebox{0.7}{\includegraphics{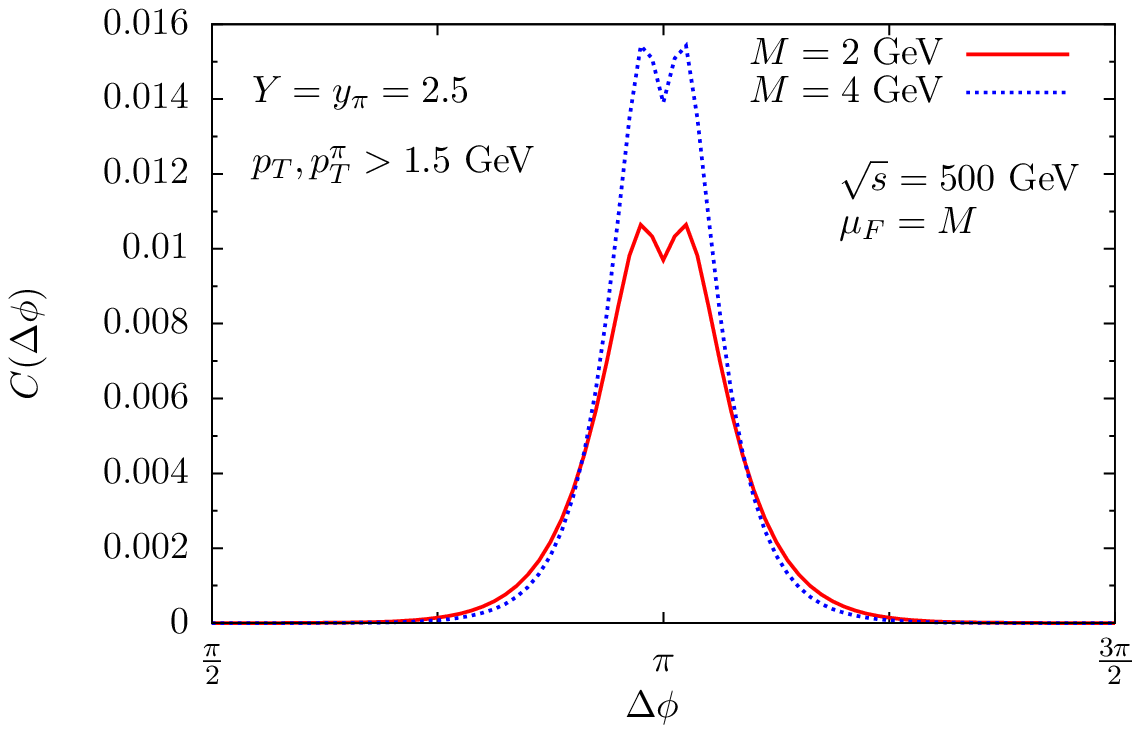}}
\caption{The correlation function $C(\Delta \phi)$ for the associated DY pair and pion production 
in $pp$ collisions at RHIC ($\sqrt{s}= 200,\, 500$ GeV).}
\label{fig:CP_pp_low}
\end{center}
\end{figure}
\normalsize

At first, we test our calculations comparing our results with those 
presented in Ref.~\cite{stasto} for $dAu$ collisions at RHIC ($\sqrt{s} = $ 200 GeV).
For this reason, following Ref.~\cite{stasto} we take the same
saturation scale $Q_{s,A}$ for a target nucleus with the mass number $A$ defined 
in terms of the corresponding scale $Q_s(x)$ for the proton target in the GBW 
parametrisation, 
${Q_s^A}^2 (x) = A^{1/3} c(b)\,Q_s^2(x)$,
where $c=c(b)$ is the profile function as a function of impact parameter $b$ 
(for central collisions we used $c=0.85$ and assume a naive GBW profile of the dipole-nucleus 
cross section following Ref.~\cite{stasto}).
Our results for forward particles $Y = y_{\pi} = 2.5$ 
and two different values of the dileption invariant mass $M$ 
are presented in the left panel of Fig.~\ref{fig:CP_rhic200} adopting 
that the factorization scale of the considered process is determined 
by the nuclear saturation scale, i.e. $\mu_F = Q_s^A$. 
%
Similarly as is in Ref.~\cite{stasto} we obtain the double-peak structure 
of $C(\Delta \phi)$ in the away-side dilepton-pion angular correlation 
with the magnitude of peaks increasing with the dilepton invariant mass 
and the width of the double peak increasing with the saturation scale. 
The normalisation of the curves turns out to be slightly 
different from that in Ref.~\cite{stasto} due to different sets 
of parton distributions and fragmentation functions used 
in our calculations, but an overall agreement is rather good.
\begin{figure}[!h]
\large
\begin{center}
\scalebox{0.7}{\includegraphics{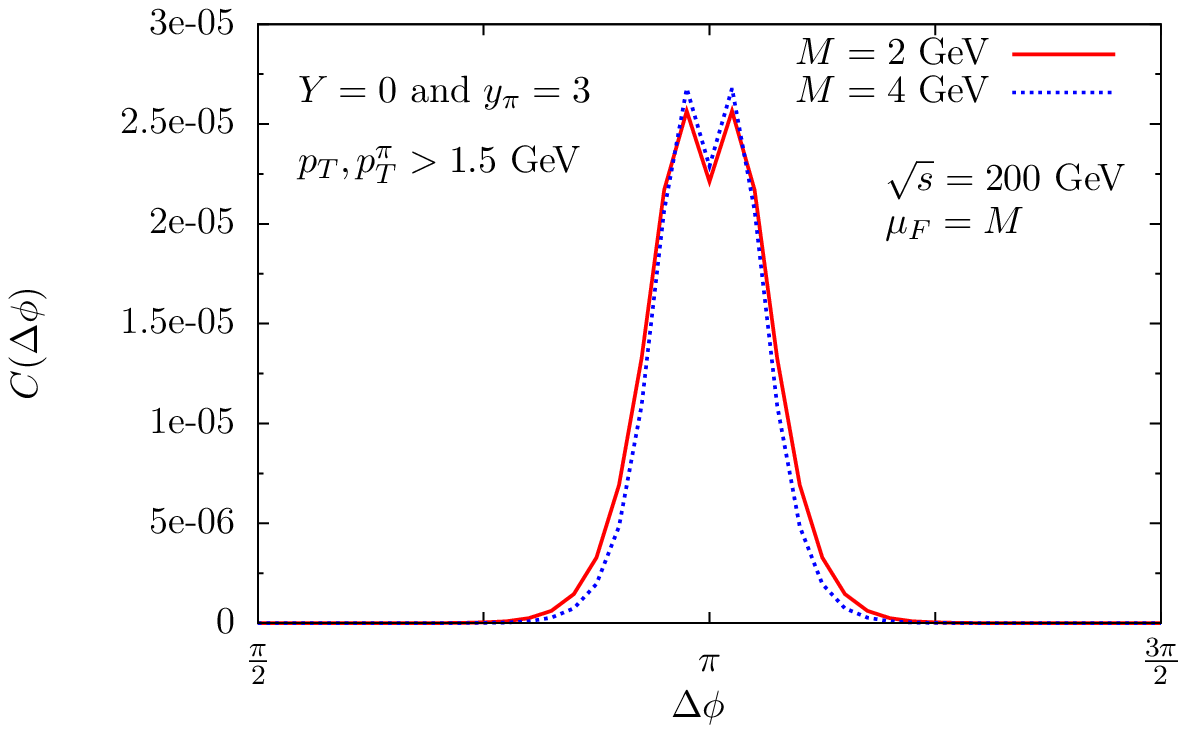}}
\scalebox{0.7}{\includegraphics{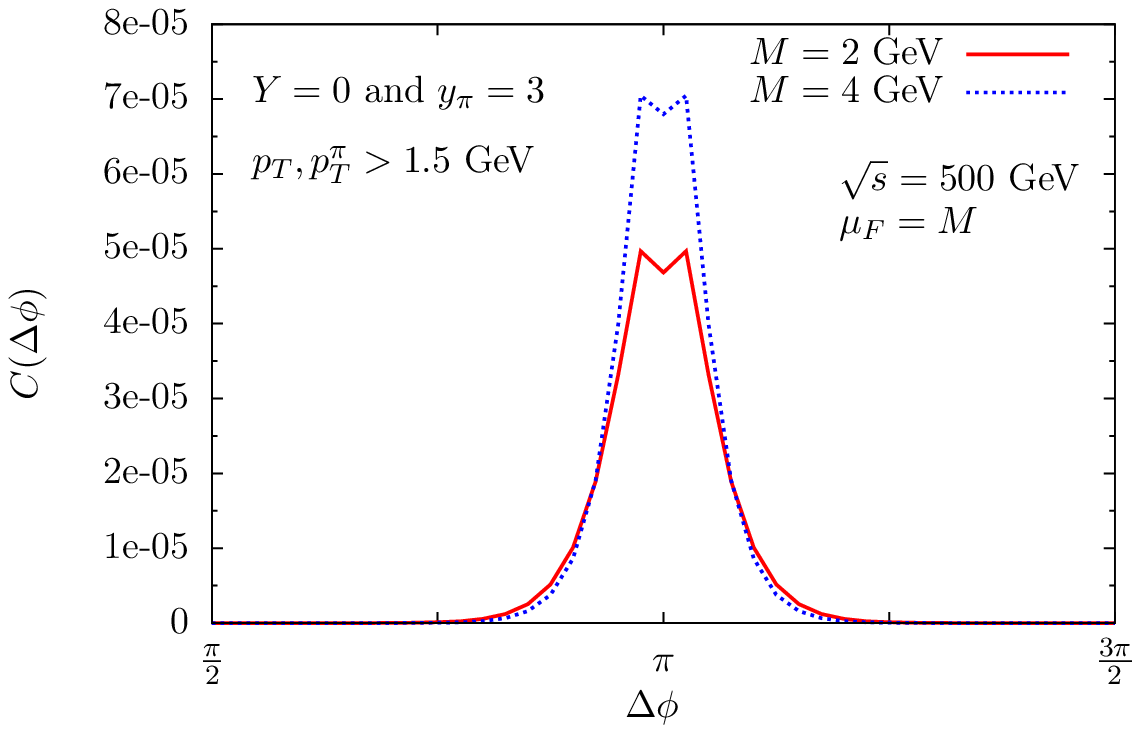}}
\caption{The correlation function $C(\Delta \phi)$ for the the associated DY pair and pion production in $pp$ collisions 
at RHIC ($\sqrt{s}= 200,\, 500$ GeV) for different values of the photon and pion rapidities.}
\label{fig:CP_mix}
\end{center}
\end{figure}
\normalsize
\begin{figure}[!h]
\large
\begin{center}
\scalebox{0.7}{\includegraphics{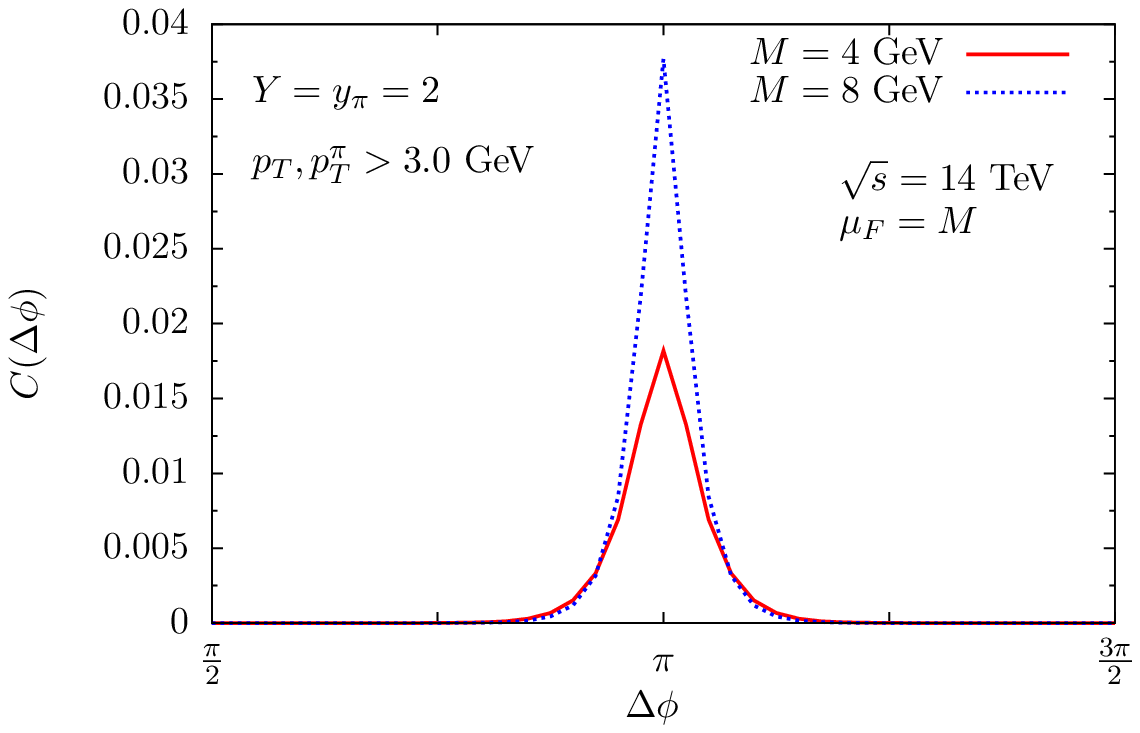}}
\scalebox{0.7}{\includegraphics{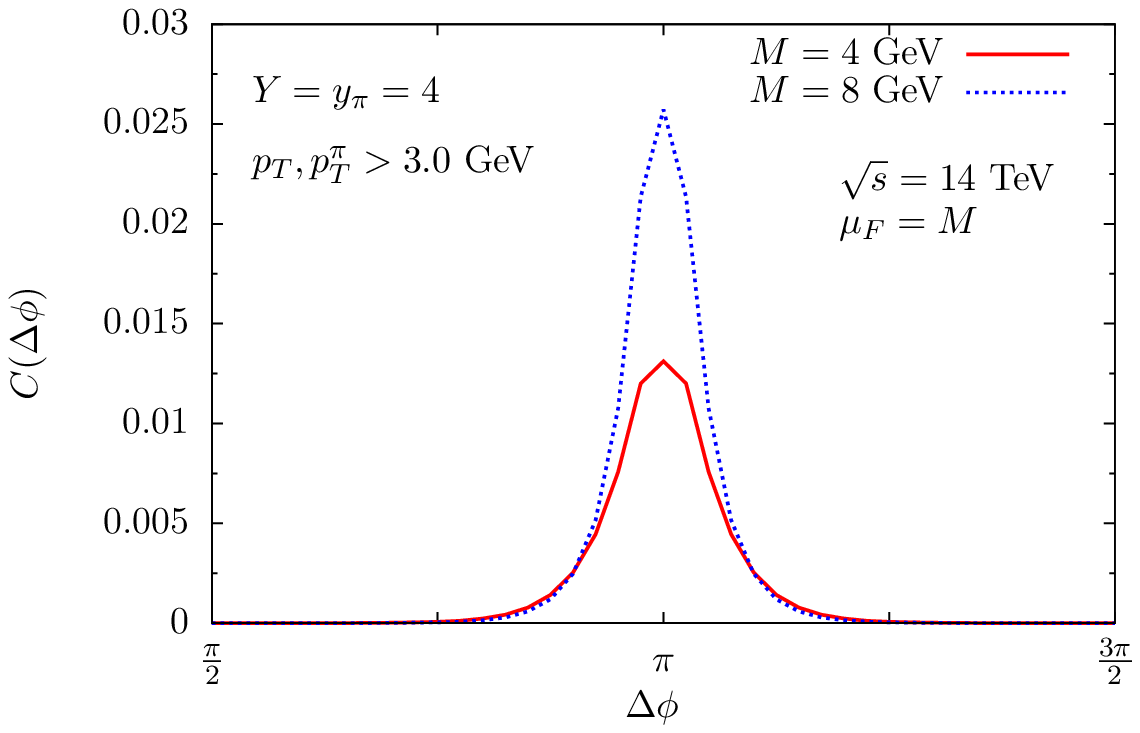}}
\scalebox{0.7}{\includegraphics{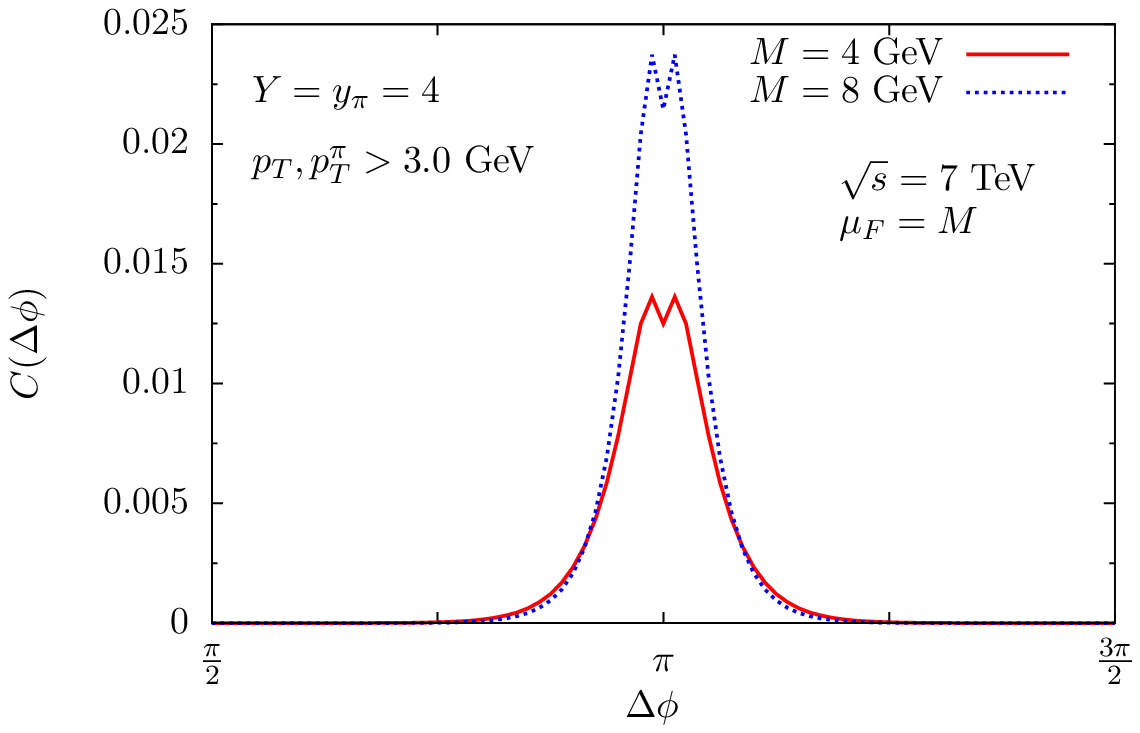}}
\scalebox{0.7}{\includegraphics{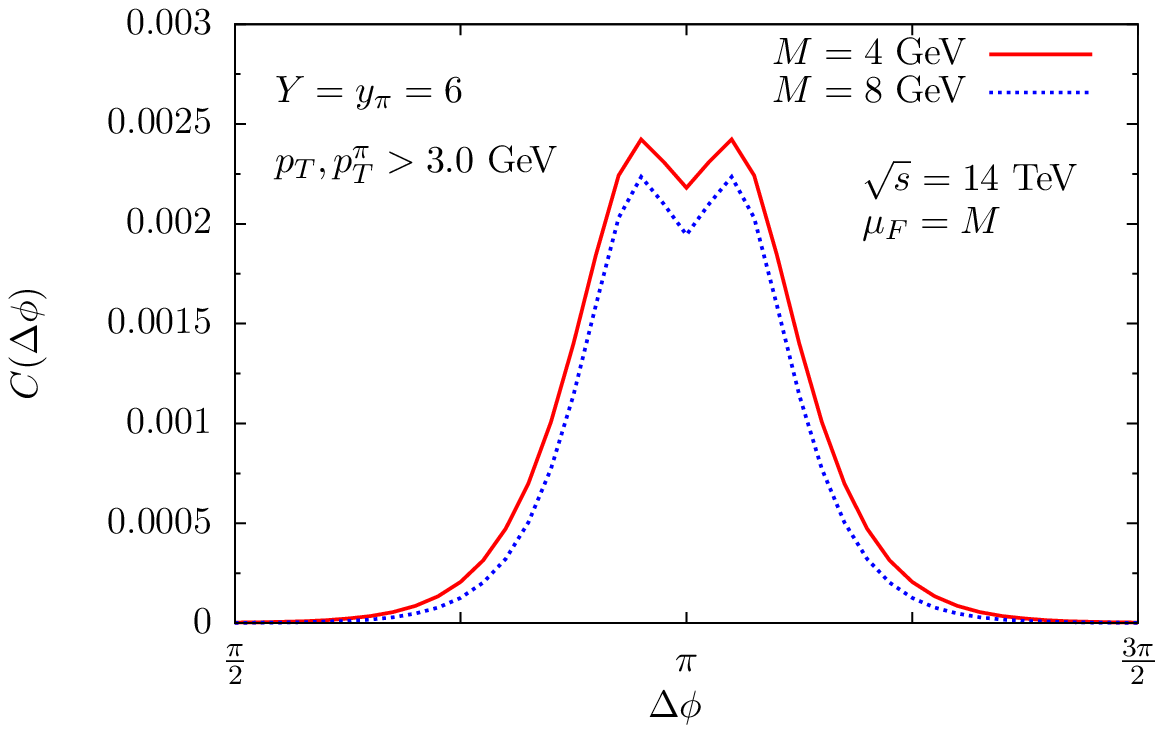}}
\caption{The correlation function $C(\Delta \phi)$ for the associated DY pair and pion production 
in $pp$ collisions at LHC ($\sqrt{s}=7,\,14$ TeV) for several values of the gauge boson and pion rapidities. }
\label{fig:CP_pp_low2}
\end{center}
\end{figure}
\normalsize
\begin{figure}[!h]
\large
\begin{center}
\scalebox{0.7}{\includegraphics{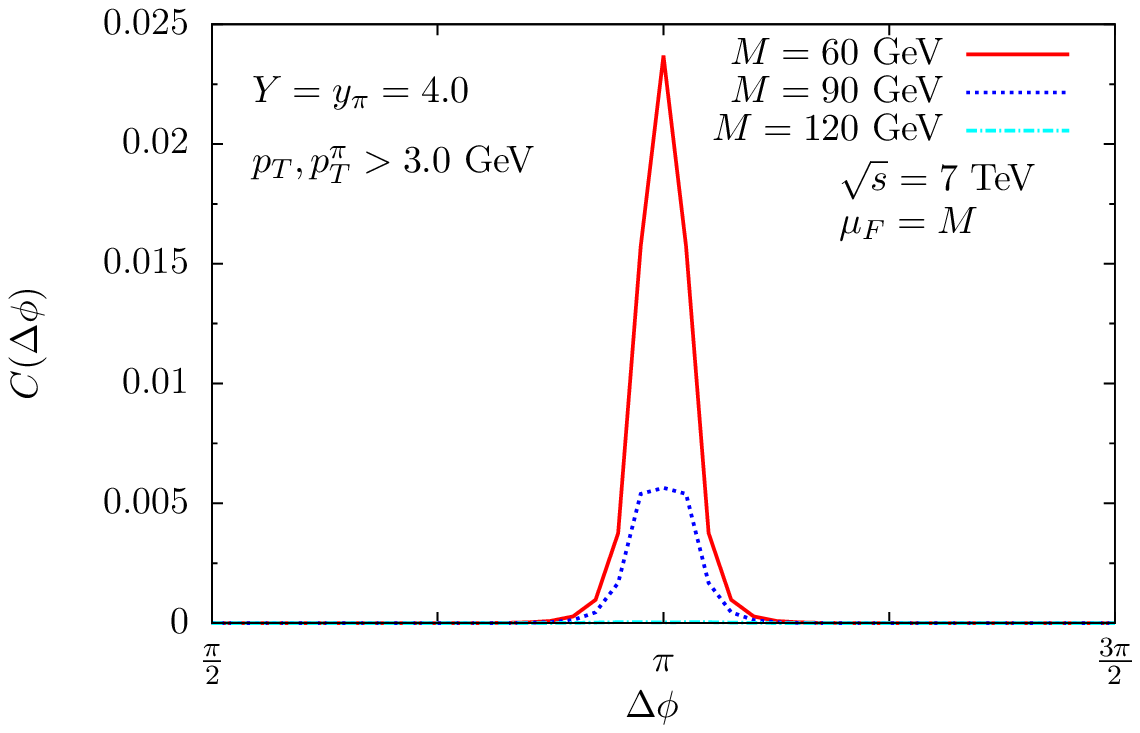}}
\scalebox{0.7}{\includegraphics{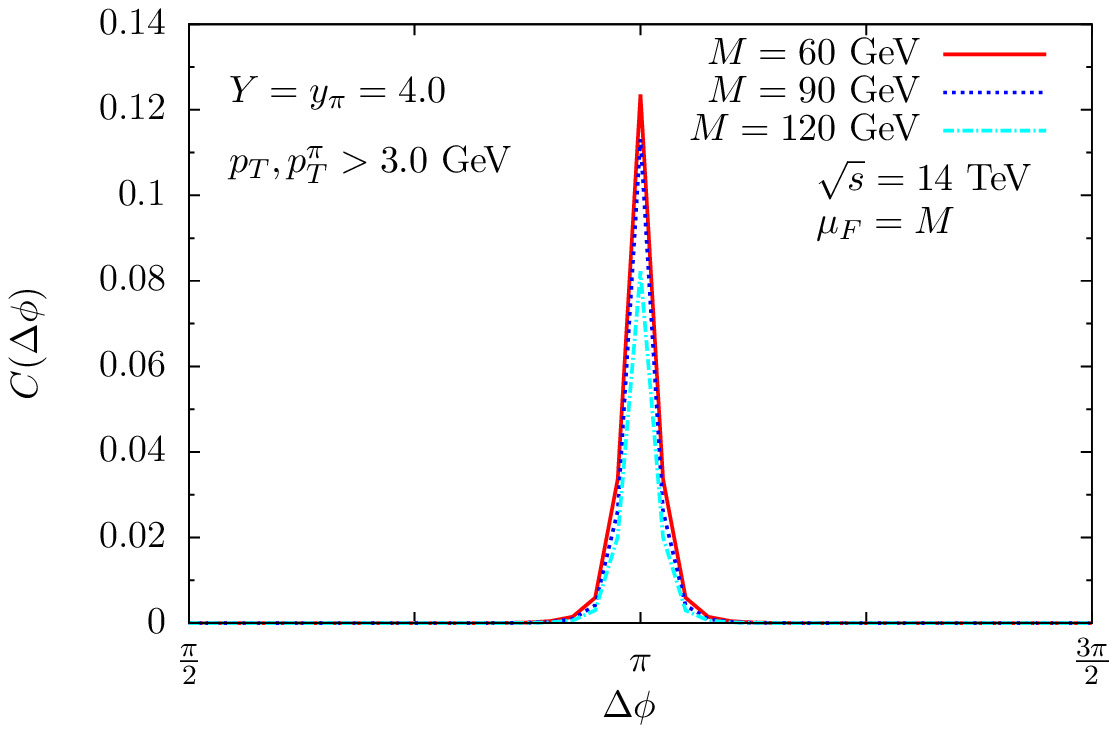}}
\scalebox{0.7}{\includegraphics{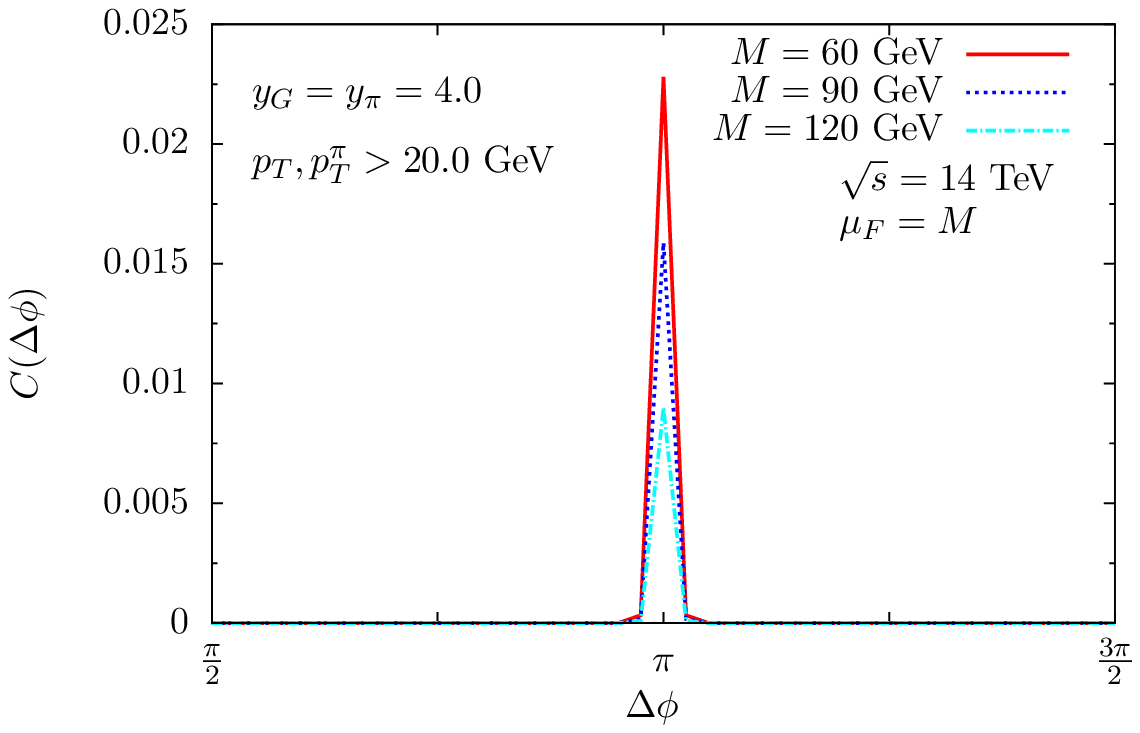}}
\caption{The correlation function $C(\Delta \phi)$ for the the associated DY pair and pion production in $pp$ collisions 
at two different LHC collision energies, $\sqrt{s}=7$ and $14$ TeV and 
two different values of the minimum cut-off $p_T^{\rm cut} =$ 3 and 20 GeV on the 
gauge boson and pion transverse momenta.}
\label{fig:CP_high}
\end{center}
\end{figure}
\normalsize

As the leading order, from the instructive point of view we test
an arbitrary choice of the factorization scale analyzing the
impact of different scale choice on results of calculation of $C(\Delta\Phi)$.  
Therefore, in the right panel of Fig.~\ref{fig:CP_rhic200} we
present prediction for the dilepton-pion correlation function
in $dAu$ collisions at RHIC for the factorization scale
$\mu_F = M$.
Such a choice of the factorization scale
is motivated by the fact that we would like to extend the formalism 
used in Ref.~\cite{stasto} also for $pp$ collisions 
and kinematical range of large invariant masses 
where $Z^0\to l\bar l$ should be included. 
Fig.~\ref{fig:CP_rhic200} clearly demonstrates that predicted double-peak structure
of the correlation function in $dAu$ collisions 
is not affected by a choice of the factorization scale. 
Consequently, the same result is expected also for $pp$ collisions as will be 
shown below.

Now let us switch on to investigation of the correlation function in $pp$ collisions. 
Figs.~\ref{fig:CP_pp_low} and \ref{fig:CP_pp_low2} 
show our predictions for the correlation function in the range of low invariant masses 
dominated by the virtual photon channel, $\gamma^* \to l\bar l$. 
In particular, Figs.~\ref{fig:CP_pp_low} and \ref{fig:CP_pp_low2} demonstrate 
that the double peak structure emerges in $pp$ collisions at both RHIC and LHC energies 
considering that the photon and pion are produced 
at forward rapidities, close to the phase space limit. 
The double peak structure of $C(\Delta\phi)$ is also predicted at RHIC energies 
for different values of the photon (central) and pion (forward) 
rapidities with corresponding results 
shown in Fig.~\ref{fig:CP_mix}. 
It is important to emphasize that such forward-central correlations
can be experimentally studied by the STAR Collaboration 
in both $pp$ and $pA$ collisions. 
In Fig.~\ref{fig:CP_pp_low2} 
we present the correlation function in $pp$ collisions at $\sqrt{s}=14$ TeV 
considering different values of the dilepton pair and pion rapidities. 
We predict that the double-peak structure of $C(\Delta\phi)$ arises only for pions 
at large forward rapidities, where the saturation scale takes values 
of the order of the dilepton invariant mass. 
Indeed, at large pion rapidities the saturation scale increases 
and becomes non-negligible compared to the typical transverse momentum of 
the back-to-back particles which induces a noticeable decorrelation between them. 
Consequently, results in Fig.~\ref{fig:CP_pp_low2} also 
demonstrate that the study of the rapidity 
dependence of the correlation function in $pp$ collisions at the LHC Run II 
can be useful to probe the onset of saturation effects.

Finally, in Fig.~\ref{fig:CP_high} we present our predictions for $C(\Delta \phi)$ 
at large values of the dilepton invariant mass imposed by the 
virtual $Z^0\to l\bar l$ channel and $Z^0/\gamma$ interference. 
In the top left and top right panels we present our results at
c.m. energy $\sqrt{s}=7$ and 14 TeV, respectively, assuming that $p_T^{\rm cut} = 3$ GeV. 
For comparison, in the bottm panel we also present results 
considering larger $p_T^{\rm cut} = 20$ GeV.
In all cases, we obtain a sharp peak for $\Delta \phi \approx \pi$, 
which is characteristic for the back-to-back kinematics of final states. 
Such a result is expected since at large invariant masses where
the effect of the intrinsic transverse momentum of gluons, which 
is of the order of the saturation scale, is negligible. 

%
%
%
%
%
\section{Summary}
\label{conc}
%
%
%
%
%

In this paper, we presented an extensive phenomenological analysis 
of the inclusive DY process $pp\rightarrow (\gamma^*/Z^0 \to l\bar l)X$ 
within the color dipole approach. At large dilepton invariant 
masses the $Z^0$ contribution to DY process becomes relevant. 
The corresponding predictions for the integrated $Z^0$ boson
production cross section as well as the dilepton invariant mass, 
rapidity and transverse momentum differential distributions 
have been compared with available data at different c.m. energies from 
Tevatron to LHC. The results were obtained employing recent 
IP-SAT and BGBK parametrisations for the dipole cross section 
accounting for the DGLAP evolution of the gluon density in the target 
nucleon. This allowed to improve an agreement with the data mainly 
at large invariant mass ranges.

Besides, we have studied correlation function $C(\Delta \phi)$ 
in azimuthal angle between the produced dilepton and a pion, which results
from a fragmentation of a projectile quark radiating the virtual gauge boson. 
The corresponding predictions has been performed at various c.m.
energies for both low and high dilepton invariant mass ranges 
as well as at different rapidities of final states. We found
a characteristic double-peak structure of the correlation function around 
$\Delta \phi \simeq \pi$ in the case of low dilepton mass, $M\sim Q_s$ 
and for pion production at large forward rapidities. 
Moreover, Figs.~\ref{fig:CP_rhic200} and \ref{fig:CP_pp_low} clearly demonstrate that
the width of a double peak arround $\Delta \phi \simeq \pi$ is strongly
correlated with the magnitude of the saturation scale $Q_s$. For this reason,
a detailed investigation of the correlation function by the future
measurements at RHIC and LHC allows to set stronger constraints on the UGDF models and 
hence to dipole model parametrisations offering thus
a possibility for more direct measurements of the saturation scale.

\section*{Acknowledgements}

E. B. is supported by CAPES and CNPq (Brazil), contract numbers 2362/13-9 
and 150674/2015-5. V. P. G. has been supported by CNPq, CAPES and FAPERGS, Brazil.
R. P. is supported by the Swedish Research Council, contract number 621-2013-428.
J. N. is partially supported by the grant 13-20841S of the Czech Science Foundation (GA\v CR),
by the Grant MSMT LG13031, by the Slovak Research and Development Agency APVV-0050-11
and by the Slovak Funding Agency, Grant 2/0020/14. M. \v{S}. is supported by the grant LG 13031 
of the Ministry of Education of the Czech Republic and by the grant 13-20841S of the Czech 
Science Foundation (GACR).

\bibliographystyle{unsrt}

\end{document}